# Fusion: It's time to color outside the lines

Wallace Manheimer

Retired from the US Naval Research Laboratory

January 29, 2024

## Abstract

There has been some good news, and some bad news in the controlled fusion community recently. The good news is that the Lawrence Livermore National Laboratory (LLNL) has recently produced a burning plasma. It succeeded on several of its shots where ~1.5-2 megajoules from its laser (National Ignition Facility, or NIF) has generated ~ 1.3-3 megajoules of fusion products. The highest ratio of fusion energy to laser energy it achieved, defined as its Q, was 1.5 at the time of this writing. While LLNL is sponsored by nuclear stockpile stewardship, this author sees a likely path from their result to fusion for energy for the world, a path using a very different laser and a very different target configuration. The bad news is that the International Tokamak Experimental Reactor (ITER) has continued to stumble on more and more delays and cost overruns, as its capital cost has mushroomed from ~\$5 billion to ~ \$25B. This paper argues that the American fusion effort, for energy for the civilian economy, should switch its emphasis not only from magnetic fusion to inertial fusion but should also take much more seriously fusion breeding. Over the next few decades, the world might well be setting up more and more thermal nuclear reactors, and these might need fuel which only fusion breeders can supply. In other words, fusion should begin to color outside the lines.

## I. Introduction

As children we all had coloring books, and the instructions our parents gave us was to color within the lines. This has been the attitude of art, painting, and sculpture from the beginning of human civilization until about 180 years ago. Look at medieval painting and ancient Greek sculptures. They all attempted to reproduce an exact image, that is they 'colored within the lines', just like we did when we were very young. Then it dawned on artists that the beauty of art could be magnified by coloring outside the lines. Monet and Renoir saw that they could enhance the beauty by blurring the lines. Van Gough saw that this could be improved still further by moving the lines around, and Picasso saw that in many works, that he did not need lines at all. We use this analogy to argue for a strategy for the controlled fusion project to 'color outside the lines'.

This article suggests an alternate course for the fusion effort. It is supported here by a rather detailed scientific argument, but one with little mathematics. It suggests that the best course of action for the development of fusion for civilian energy is to 'color outside the lines'. In other words, the paper suggests strategies for fusion which generally have not been used in the bureaucratic world that the American fusion project lives in. Briefly, it suggests that the US Department of Energy set up a lab (possibly in one of the existing DoE labs, possibly an entirely new lab) which will work on direct drive laser fusion, most likely with an excimer laser, for energy the civilian economy. This is motivated in large degree both by the tremendous success of the recent laser fusion experiments of the Lawrence Livermore National Lab (LLNL) (Zylstra A, Kritcher, Abu-Shawareb, Zylstra B, Seminar, Divol), as well as by what, at this point look like nearly overwhelming obstacles that magnetic fusion must overcome to succeed. If this new lab could replicate the LLNL result in a direct drive configuration, with a far ultraviolet laser (193nm), it would be an event of incredible importance. It would then be very likely that there exists a path from that accomplishment to fusion energy for the world. This new lab would cooperate with, and possibly somewhat compete with LLNL, which is working on laser fusion, their goal being stockpile stewardship and national nuclear security. Since the goals of the two labs would be quite different, it seems likely to this author that cooperation would be maximized, and competition would be minimized.

Of course, I hardly think that the US Department of energy would embark on such change on my word alone. My hope is to convince a portion of the fusion and larger scientific establishment to discuss and perhaps support such a change. Accordingly, recently the author has made that argument in a variety of different media, including a book (Manheimer 2023 A), an essay in the American Physical Society journal Forum on Physics and Society (Manheimer 2024 A), a podcast (Manheimer 2024 B), and a seminar at NRL (Manheimer 2023 B). The author hopes that this article in arXiv, and possibly also later in an archived scientific journal completes such a quintet. He hopes this makes the case that this is an essential discussion to have, and that this discussion is taken seriously.

At this point is appropriate to comment that at least in the United States, the government supported fusion program is hardly the only game in town. Many 'fusion startups' funded by many private dollars have recently appeared on the scene (Clynes, Kramer B, helion, tae,

cfs…..). They have gotten a tremendous amount of publicity, promising commercial fusion hooked up to the grid in a decade or a bit more. Recently a White House conference took place on these (McCarthy). The theme was that fusion, beyond the national lab approach, will succeed sooner.

Their claim was that they can put fusion on the line just in time to avoid a 'climate crisis'. This paper does not ignore such an 'elephant in the room' but does dismiss it. Furthermore, it denies the threat of an imminent climate crisis, as have many first rate scientists, economists, engineers and the like ($CO_2$, Wijngaarden, Lamb, Christy, Moore, Watts, Manheimer 2023A, Lomborg, Epstein, Koonin, Lindzen, Spencer, Manheimer 2022A). Lamb, in fact wrote the classic textbook on climate, and showed that there are many influences on climate besides $CO_2$. While he does not neglect the effect of $CO_2$, he does not mention it until page 330. He calculated that with an increased in $CO_2$ from 400 to 800ppm, the temperature would increase by ~1°C and increase much more slowly with further added $CO_2$. His result is consistent with some of the most detailed recent calculations (Wijngaarden)

There are arguments, and frankly, propaganda, in the mass media (Gelles A, Gelles B, Todd), social media (Google), and statements of prestigious scientific societies (aps, ametsoc, acs) that there is a rapidly on rushing climate crisis unless the world quickly ends its use of carbon-based fuel. This assertion is simply false. It is more than discouraging to see these prestigious scientific societies snap at the bait, hook, line, and sinker, and doing so without performing even minimal due diligence to consider opposing views, such as in Lamb's classic textbook. They will almost certainly have to answer some tough questions about their actions in the not-so-distant future (Manheimer 2023 D). There is voluminous evidence proving the absence of a climate crisis (see above), including signed statements by thousands of very knowledgeable scientists asserting that there is no climate crisis now or in the foreseeable future ($CO_2$, petitionproject, clintel).

If there is any agreement among the most qualified scientists, it is that $CO_2$ in the atmosphere, is *not* a pollutant. Note that $CO_2$ is an atmospheric gas that is absolutely essential for life. If the atmospheric $CO_2$ level dropped to zero (or more accurately, below ~150 ppm), there would soon be no life on this planet. This is undeniable. In its geological history, levels of atmospheric CO2 have been all over the place, with the preindustrial level of ~ 280 ppm being nearly the lowest level (Davis). In fact, let those asserting that atmospheric $CO_2$ is a dangerous pollutant answer this simple question which they do not seem to have ever considered: What is the optimum level of $CO_2$ in the atmosphere and why?

Regarding the claim of fusion online in a decade or so, there is a great deal of literature (Stiffler, Jassby 2019, Jassby 2021, Steiner, Manheimer 2023A, Lampe, youtube, Buttery) making the case that the obstacles between where fusion is now and where it needs to be, are much too great to overcome in about a decade. The authors of these skeptical papers are for the most part experienced fusion scientists, but who are currently retired, are financially independent, and have no bosses or sponsors they must please. Hence this paper favors the more standard government sponsored approach for fusion development. The goal is simply too far away for private investors to risk their capital on it; and there is no necessity for fusion's rapid development. Remember, the tortoise always beats the hare.

Getting back to this US Department of Energy new lab, it would not only investigate pure fusion but would also examine fusion breeding. That is using fusion neutrons to breed fissile material for use in thermal nuclear reactors. As we will see, fusion breeding has advantages which neither fast neutron breeder reactors, nor thermal thorium breeder reactors have. Specifically, unlike the fast neutron or thermal thorium breeder, a fusion breeder can fuel many thermal nuclear reactors which in the next few decades might have difficulty obtaining fuel. But fusion breeding has been the ugly duckling of both the fusion and fission world, often disparaged by such ignorant descriptions as 'combining the worst aspects of both fusion and fission'. This work, and others hope to turn the ugly duckling into a beautiful swan.

In the most likely case that no new US government support for this fusion effort can be found, this means reprograming hundreds of millions of $$ from magnetic fusion to inertial fusion in the US. Obviously, many entrenched scientific and bureaucratic interests would fight this tooth and nail (i.e. bite and scratch). Overcoming this obstacle would be like crossing a quicksand bog a mile wide and a mile deep. However, this is an important and necessary switch. This paper gives scientific arguments and justifications for this advised major switch in strategy. In other words, fusion must begin to color outside the lines.

This author has argued for fusion breeding for decades in both earlier work (Manheimer 1999, Manheimer 2009, Moir 2013) and more recently in a series of detailed articles published open access in prestigious, well-established journals (Manheimer 2014, Manheimer 2018, Manheimer 2020 A and B). Anyone would consider these journals to be first class scientific journals. Also, a shorter version for a less technical audience was published in the Forum on Physics and Society (Manheimer 2021 A). However, these articles, for the most part greatly deemphasized the role of laser fusion. When they were written, laser fusion did not look to the author like a contender, so despite all the problems magnetic fusion faced, these articles mostly concentrated on tokamaks. The recent LLNL experiment changed that.

In addition, this author's most recent work on breeding, including laser fusion, has been published in an Indian journal (Manheimer 2022 B). After being accepted, rather enthusiastically, by 3 reviewers for an article in a special issue of Fusion Science and Technology on exotic uses of fusion, it was finally rejected by the publisher. This author's guess it that the work was 'cancelled' for its skepticism on wind, solar and batteries. Its appendix relates this experience. This author has the copyright for these open access articles. It is a copyright which allows anyone the freedom to copy the work for any reason, as long as the work is properly cited. Hence portions of this article have been taken directly from these open access publications by the author. Of course, there is also a great deal of new material in this article.

In Section II, we first define the overall goal, namely, to achieve OECD scale power world-wide, that is about 35-40 terawatts (TW), or about 4 kilowatts (kW) per capita by midcentury or as soon after that as possible. This means that the world must make a large-scale transition to nuclear power, and especially to breeding. We show here that fusion breeding, if we can pull it off, has many advantages over other breeding possibilities. In Section III, we discuss fusion research within the lines, namely tokamaks (and to a lesser degree stellarators) for magnetic fusion energy, and laser fusion for nuclear stockpile stewardship. Section IV discusses why

these appear to be inadequate for the future development of fusion energy. Section V discusses how coloring outside the lines, by switching on a large scale to an excimer laser driven fusion may be the salvation. In Section VI we discuss fusion breeding, Section VII gives several digressions which seem important to the goal, and Section VIII discusses bureaucratic aspects of the switch. Section IX discusses 'the energy park', a proposed element to a nuclear energy infrastructure which is economic, environmentally viable, and has little or no proliferation risk. Finally, Section X draws conclusions.

Section II:  Why Fusion?

Why has the world supported fusion for over 60 years, a goal which seems to recede at least one year per year?

The reason is that modern civilization needs energy.  Before fossil fuel became widely used, this energy was provided by people and animals.  Because this constituted so little energy, civilization had been a thin veneer atop a vast mountain of human squalor and misery, a veneer maintained by such institutions as slavery, colonialism and tyranny.  This veneer allowed the ancient Egyptians to build gigantic pyramids, and  allowed the ancient Greeks to build magnificant temples and statues, while their much larger underclasses and slaves lived in misery and squalor.

The desire for a better life style among these slaves and underclasses is universal, and probably is universal among all well meaning people.  In Hebrew, it is called Tikum Olam, or repairing the world.  But to do this takes much more energy than what was available in the ancient and more recent historical world.

Fossil and nuclear fuel has extended the benefits of modern civilization to billions, but its job, in this respect is not yet complete.  There are still billions on earth who derive little benefit from this power source, and billions more who derive hardly any. To spread the benefits of modern civilization to  the entire human family would require much more energy, as well as newer sources of energy.

One excellent source of these statistics is the yearly publications by (BP).   Taken from their 2019 issue are their graphs of the sources of energy, the energy use in various parts of the world, and by end use shown in Fig (1).

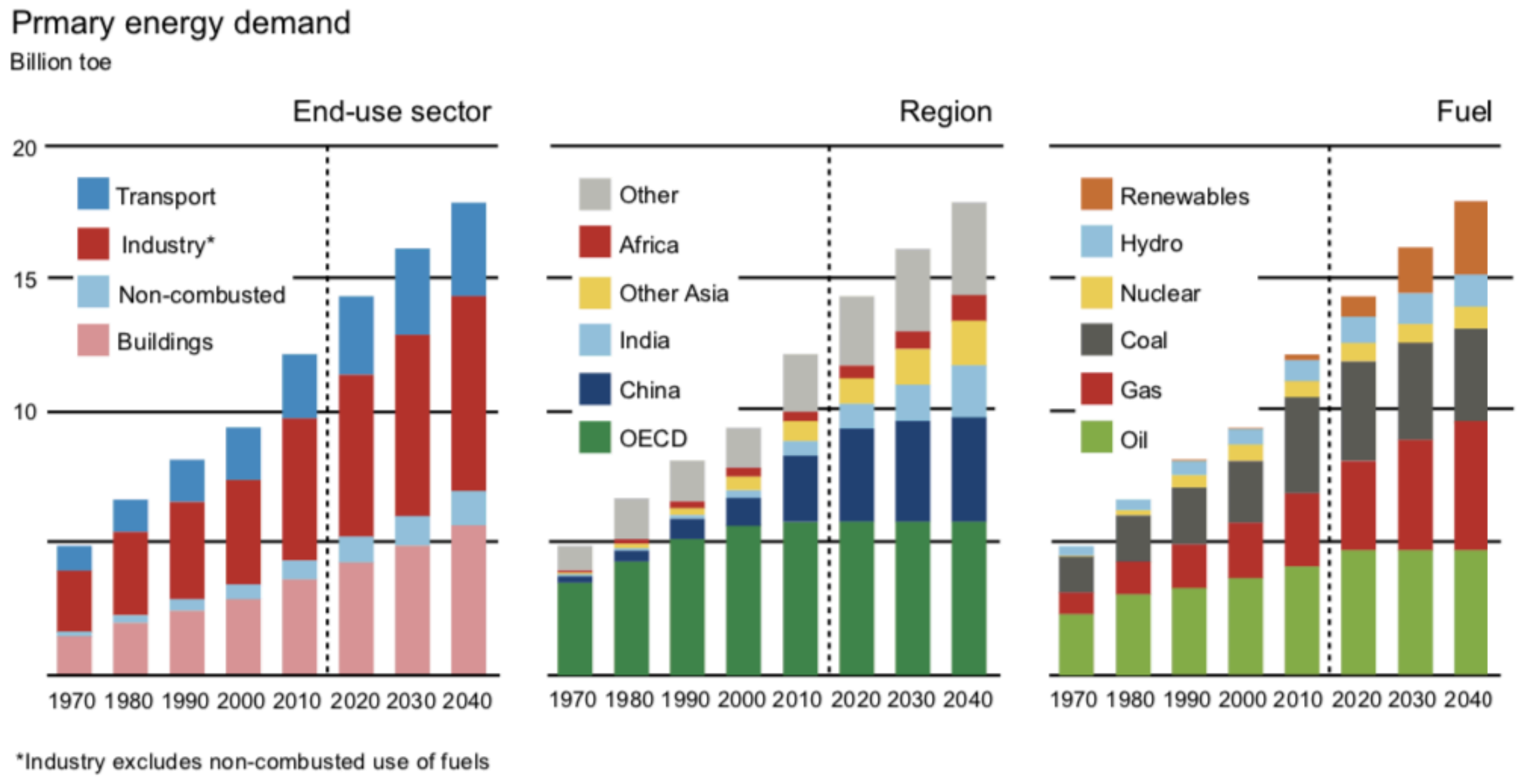

Figure 1: BP's three graphs of energy demand by end use sector, region and fuel. The units are billions of tons of oil per year. Since this is an unusual unit for people not in the oil industry, we use terawatts (TW), where one terawatt is approximately one billion tons of oil per year. Taken from (BP).

To the left of their vertical dashed line in Fig. (1) is the historical record. To the right are BP's extrapolations for the future. Plagiarizing a bit from the American Declaration of Independence, the author holds this truth to be self-evident, namely that it must be the goal of the world to bring the entire world up to OECD standards, soon, say by mid-century. As is apparent from the graph, the world today uses about 14 terawatts (TW). However the energy use is very unequal. The 1.2 billion people in the economically more advanced OECD countries use ~ 6 TW, or ~ 5 kilowatt (kW) per capita. The other 7 billion or so people living on the planet share 8 TW, or use ~ 1 kW per capita. The world's goal certainly must be to bring the rest of the world up to OECD standards of living as quickly as possible. By mid century the world population is expected to level off at about 10B, meaning that at current OECD power use, the world would need as much as 35-40 TW assuming that energy efficiency increases by ~ 30% by then (Hoffert 1998 ).

Whether the concern is exhausting fossil fuel (we can use it for quite a while but will exhaust it in 1/3 the time at 35TW as at 12), or is knowing that solar and wind cannot do the job (Mills 2019 & 2020, Manheimer (2021B & 2022C) or knowing that pure fusion cannot do the job, at least in this century if ever (Reinders), these lead to one and only one conclusion. Nuclear power must play an important role, both in any final sustainable role, and on the way there.

Very briefly, nuclear power is based on splitting atoms as shown in Fig (2). If a neutron strikes a $^{235}$U nucleus, it splits it into 2 fragments, which have an energy of ~ 215 MeV, and produces an average of ~ 2.4 additional neutrons. These neutrons can cause a chain reaction. However, only 0.7% of the world's total uranium resource is $^{235}$U. Actinides (i.e. nuclei with atomic number 92 or greater), such as $^{233}$U or $^{239}$Pu, with odd atomic weight are also possible fuels. However, these fuels do not exist in nature, but most be bred from bombarding $^{232}$Th or $^{238}$U with neutrons. The difference between these odd atomic number, and the even atomic number nuclei, like $^{238}$U and $^{240}$Pu, is that odd atomic number actinides have much greater fission cross section at low energy, than do either odd or even nuclei at the 2 MeV energy at which they are produced in the nuclear reaction. Hence and important part of all existing nuclear reactors is a means of slowing down the neutron to a point where its reaction cross section is sufficiently large to continue the chain reaction. For this reason, these reactors are called thermal nuclear reactors.

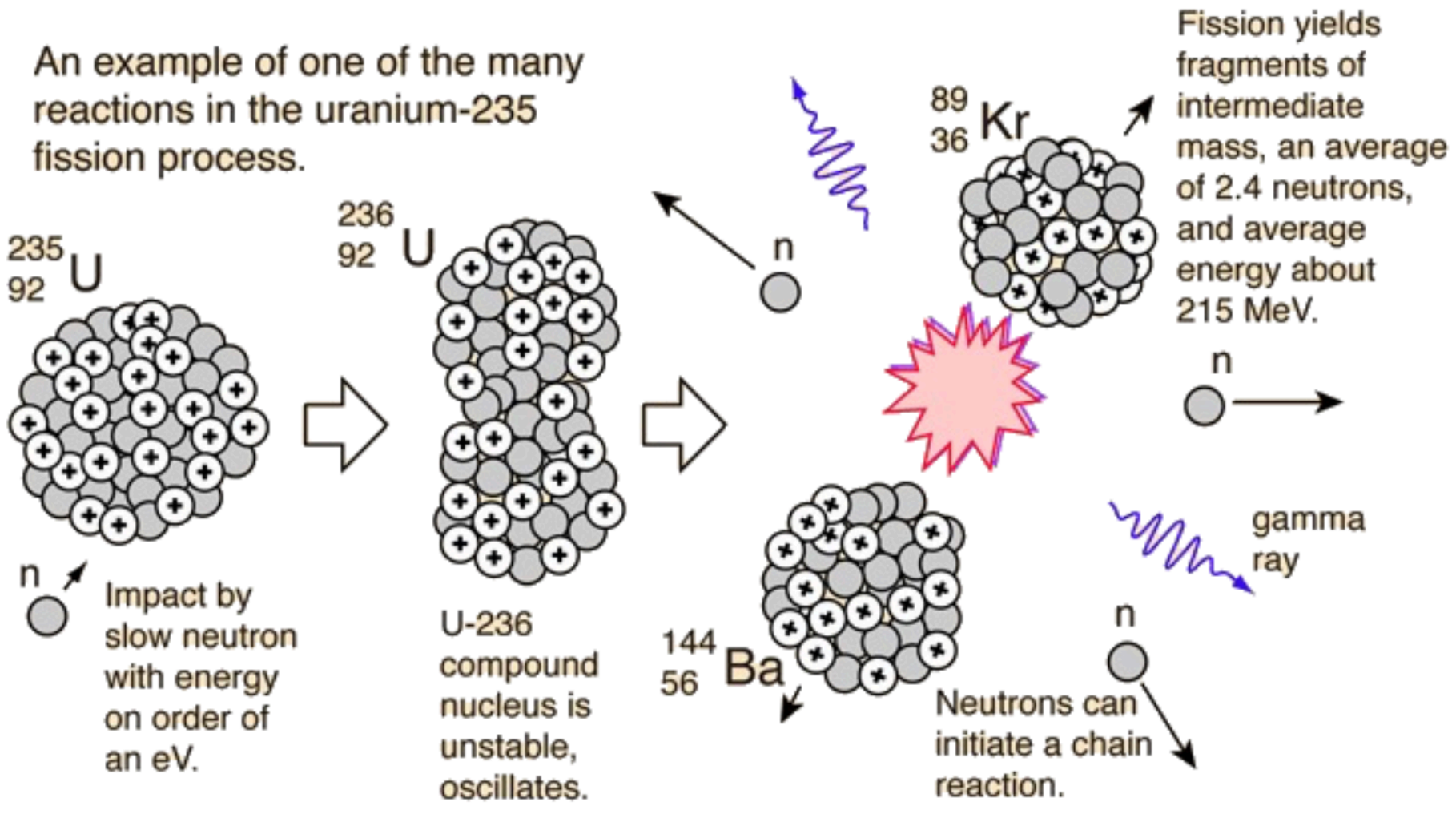


Figure 2: A graphic of the elements of the nuclear reaction, of the type which all thermal reactors use.

Right now, there are ~400 nuclear power plants in the world (~100 in the US). Most of these are light water reactors (LWR's), which use water as the coolant, and use collisions of energetic neutrons with the hydrogen in the water molecule to slow down the neutrons.

Each of these reactors generate ~ 3 gigawatts (GWth) of thermal power and ~1GWe of electric power. It is initially fueled with about 25 tons of uranium, about 1 ton of $^{235}U$, the fissile material, and ~24 tons of $^{238}U$, the fertile material. At the end of a year, the waste fuel is discharged, now containing still ~24 tons of $^{238}U$, 0.8 ton of highly radioactive intermediate atomic number (Z) fission fragments, and 0.2 tons of $^{239}Pu$ and other higher Z actinides. Note that the reactor not only burns the $^{235}U$, but also converts some of the $^{238}U$ to $^{239}Pu$. Some of this plutonium is also burned in the reactor as it is produced, but not all of it; the rest is expelled when the reactor is refueled. Hence as a rough general rule, we may think of a ton of $^{235}U$ as generating 1 GWe for a year.

Let us think of a sustainable future for all mankind as one that increases nuclear power to ~ 20-25TW (i.e. ~6.5-8.5TWe) worldwide by midcentury, and reducing fossil fuel slightly to ~10TW, so it will last at least as long as current estimates and be available to support various chemical industries. At the current rate, this would increase $CO_2$ into the atmosphere by ~ 2ppm/year, to an atmospheric concentration to ~800 ppm in ~ 200 years. We do not regard the use of fossil fuel, at 10 TW well into the future, as causing any extreme, or even any minor planetary calamity, at least in the next 200 years. There is solid scientific work supporting this assertion. Hence, let us also think of increasing hydroelectric power to 2-3 TW, and other forms, perhaps garbage to energy, windmills… to 1-2 TW. Furthermore, it does not regard solar and wind as viable large scale power sources, but perhaps they could play a minor role in niche markets.

This then would obviously require something of a crash program in expanding nuclear power over the next few decades. There is every reason to think this possible technically, although perhaps not politically. At least in the United States, regulations, lawsuits, protest marches, bureaucratic delays, environmental impact statements done and redone numerous times, NIMBY, BANANA,… have all thrown sand in the gears of nuclear power. These could be the biggest problem it faces. Even if the nuclear company is successful, typically 20 years are wasted as it strangles in bureaucratic red tape and court cases, enormously increasing the price of nuclear power. Delay means $$$, time is money. Regulation reform is the American, and perhaps the worldwide nuclear industry's biggest battle right now.

Yet even if the nuclear industry solves its regulation problem, over the long term it faces a much bigger problem on the physics and technical side. Fissile $^{235}U$ comprises only 0.7% of the uranium resource. Supplies of mined $^{235}U$ are limited, almost certainly much less than the reserve of fossil fuel. One rather pessimistic estimate is that the energy resource of mined uranium is about 60-300 Terawatt years (Hoffert 2002). Other estimates are higher (Freidberg 2015), but no estimate is high enough, that if it were correct, there would be enough uranium to sustainably supply the world's thermal nuclear reactors with 20-30TWth (i.e. ~7-10TWe).

Over the years, this author has been in contact with several experts in the field of nuclear reactors. One of these was Daniel Meneley (deceased 2018), who was once in charge of the Canadian program and worked on both the heavy water moderated CANDU (Canada Deuterium Uranium) reactor, and the Integral Fast Reactor (IFR), built by Argonne National Laboratory in the US. In 2006, he asserted in 2 separate emails (Meneley)

I've nearly finished prepping my talk for the CNS on June 13th (2006) -- from what I can see now, we will need A LOT of fissile isotopes if we want to fill in the petroleum-energy deficit that is coming upon us. Breeders cannot do it -- your competition will be enrichment of expensive uranium, electro-breeding. Good luck.

And:

We (I'm on the Executive of the Environmental Sciences Division of ANS) held a "Sustainable Nuclear" double session at the ANS Annual in Reno a couple of weeks ago. I have copies of all the presentations. ........... The result was an interesting mixture of "we have lots", just put the price up and we'll deliver (we've heard the same from Saudi recently) and "better be sure you have a long-term fuel supply contract before you build a new thermal reactor".

So let's imagine that the world has gone largely to nuclear power, as suggested here. If so, it is entirely possible that 30-50 years from now, we will be stuck with thousands of reactors, but fuel for them will become very expensive, or difficult to get, or unavailable. Uranium from the oceans cannot save them, there is only 1.8 MJ of $^{235}U$ in each metric ton of sea water (Hoffert 2002, Guidez, Manheimer 2023A). It will take more energy than that to collect and process it.

What then? One might think that fast neutron breeders (Garwin) could be the solution. It is true, that they, and thermal thorium breeders (Freeman), could be the be architecture of sustainable energy system, but they have no ability to fuel thousands of 'stranded thermal' reactors which are out of fuel.  As we will see, the only breeder which has the capability of doing this is a fusion breeder.

Breeding means taking a material which exists in nature like, $^{238}U$ or $^{232}Th$, called fertile materials, and bombarding it with neutrons to make fissile materials, like $^{239}Pu$ or $^{233}U$, which do not exist in nature.  However, because these have an odd atomic weight, they are fine as fuel for thermal fission reactors such a light water reactor (LWR).

Not only is the reaction cross section much greater for a thermal neutron reactor, but the thermal reactor designer has a wide choice of coolants (e.g. water or air), instead of only liquid sodium or lead, which must be used in a fast neutron reactor. Figure (3) is a plot of the fission and neutron absorption cross sections as a function of neutron energy for $^{235}U$ and $^{238}U$ (National).

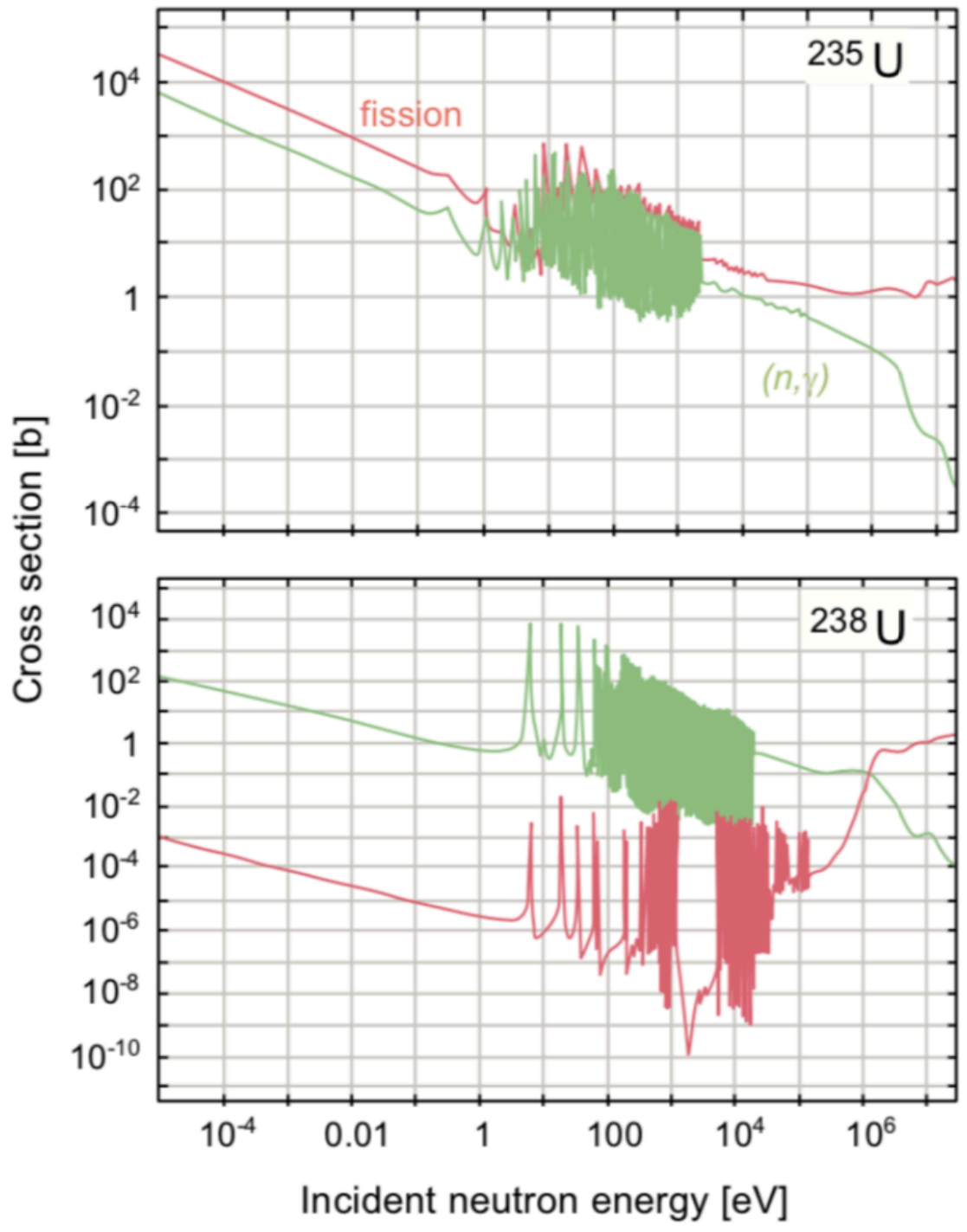

Figure 3 : The fission and neutron absorption cross section in barns (1 barn is $10^{-24}$ cm) for $^{235}U$ and $^{238}U$ as a function of the energy of the incident neutron. The cross sections look about the same for all fertile and fissile nuclei, depending on whether their atomic number is odd or even.
The red curves are the fission cross sections, and the green, are the neutron absorption cross sections.

Fission breeders, of course can, and have been developed.  France also had its Super Pheenex breeder hooked up to its grid for a while until the greens' constant protests succeeded in having it decommissioned.  It finally worked, but it took years to iron out all the bugs, principally difficulties with dealing with liquid sodium in the quantity necessary to cool the reactor (Garwin). Other countries, especially Russia and India are also taking these reactors very seriously. Russia already has two fast neutron reactors, their BN 600 and BN 800 attached to their grid. (The Russian word for fast is bistro.) The United States had developed a 60 MW breeder at Argonne National Labs called the integral fast reactor (IFR) which ran successfully for years (Beynon, Hannum). It could run either as a breeder or burner. Even as a burner, it has an advantage, that despite its high cost, it can burn any actinide equally, whether it has an even or odd atomic number. In 1994, work on it was abandoned, largely at the instigation of Senator Kerry, who saw it as a proliferation threat.

The fast neutron breeder reactor has an unavoidable drawback. It is not a very prolific producer of fissile material. Another contact was with George Stanford, a nuclear engineer and physicist who was a key member of the design team for the IFR. In 2006, he wrote (Stanford):

Fissile material will be at a premium in 4 or 5 decades.....I think the role for fusion is the one you propose, namely as a breeder of fissile material if the time comes when the maximum IFR breeding rate is insufficient to meet demand.

The reason a fission breeder is it is unlikely to meet demand, as both Daniel Meneley and George Stanford stipulated is very simple. The fast neutron reaction does produce more neutrons than thermal neutrons, but not that many more. Let us think of a thermal reaction as producing 2.4 neutrons, and a fast neutron reaction a producing 3. Of these 3, one continues the chain reaction, one replaces the nucleus that produced the original reaction, half a neutron is lost due to some loss mechanism, leaving one neutron for every two reactions which can be used for other purposes. Let's assume that this other purpose is to produce a fissile $^{239}$Pu, or $^{233}$U from a fertile $^{238}$U or Th. This can be used as fuel for another thermal reactor. However, it would take two fast neutron breeders to fuel a single thermal reactor of equal power. Figure (4) illustrates this.

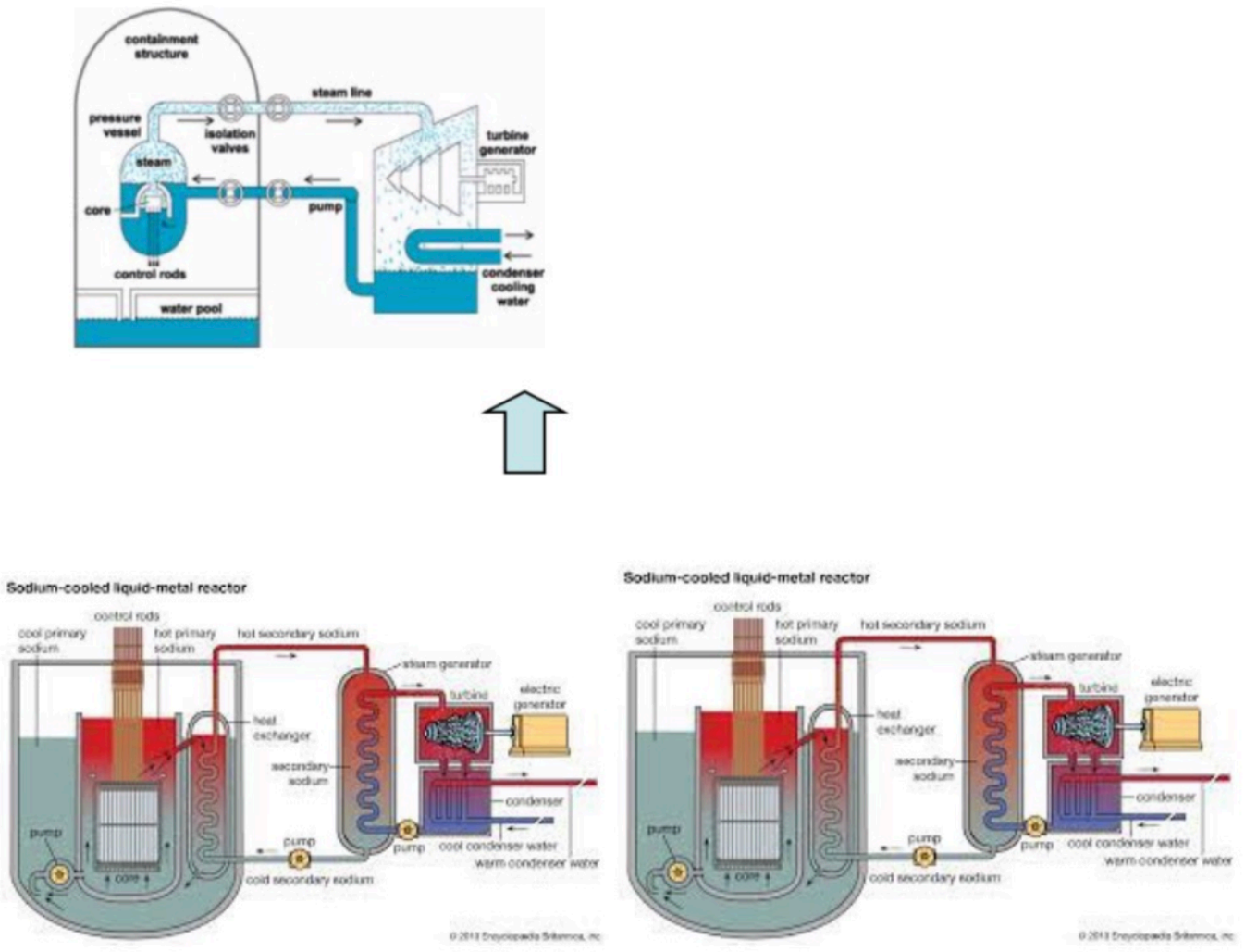

Figure 4: It takes two fast neutron breeders, at maximum breeding rate, to fuel a single thermal reactor of equal power.

Another breeding option is a thermal thorium breeder (Freeman). When a thorium nucleus absorbs a neutron, it becomes unstable, and quickly ejects an electron to move one level up in the periodic table to become protactinium (Pa), which is also unstable. It too ejects an electron and becomes $^{233}$U, a perfectly good fissile material which fuels the reactor. Basically, the thorium breeder substitutes thorium for $^{238}$U as the fertile material. The neutron economy is such that it can refuel itself from the thorium. There is plenty of available thorium, so it is a sustainable nuclear architecture. However, the thermal thorium breeder can only fuel itself, it has no capability of fueling any other thermal reactor.

As we will see, only a fusion breeder can fuel many thermal reactors. Very briefly this takes advantage of the fusion neutron's high energy to produce additional neutrons by a process called spallation. Then these neutrons can collide with either fertile $^{238}$U to ultimately generate fissile $^{239}$Pu, or collide with fertile $^{232}$Th to produce fissile $^{233}$U. Either of these is a perfectly good fuel for a thermal nuclear reactor. As we would like to eliminate plutonium to the extent possible, we consider here only the thorium/uranium233 route. This author is not aware of any detailed plans for such a fusion breeder, but a simple calculation, appearing later in this paper, suggest that a fusion breeder can fuel at least 5, and possibly as many as 10 thermal reactors. This is illustrated in Figure (5).

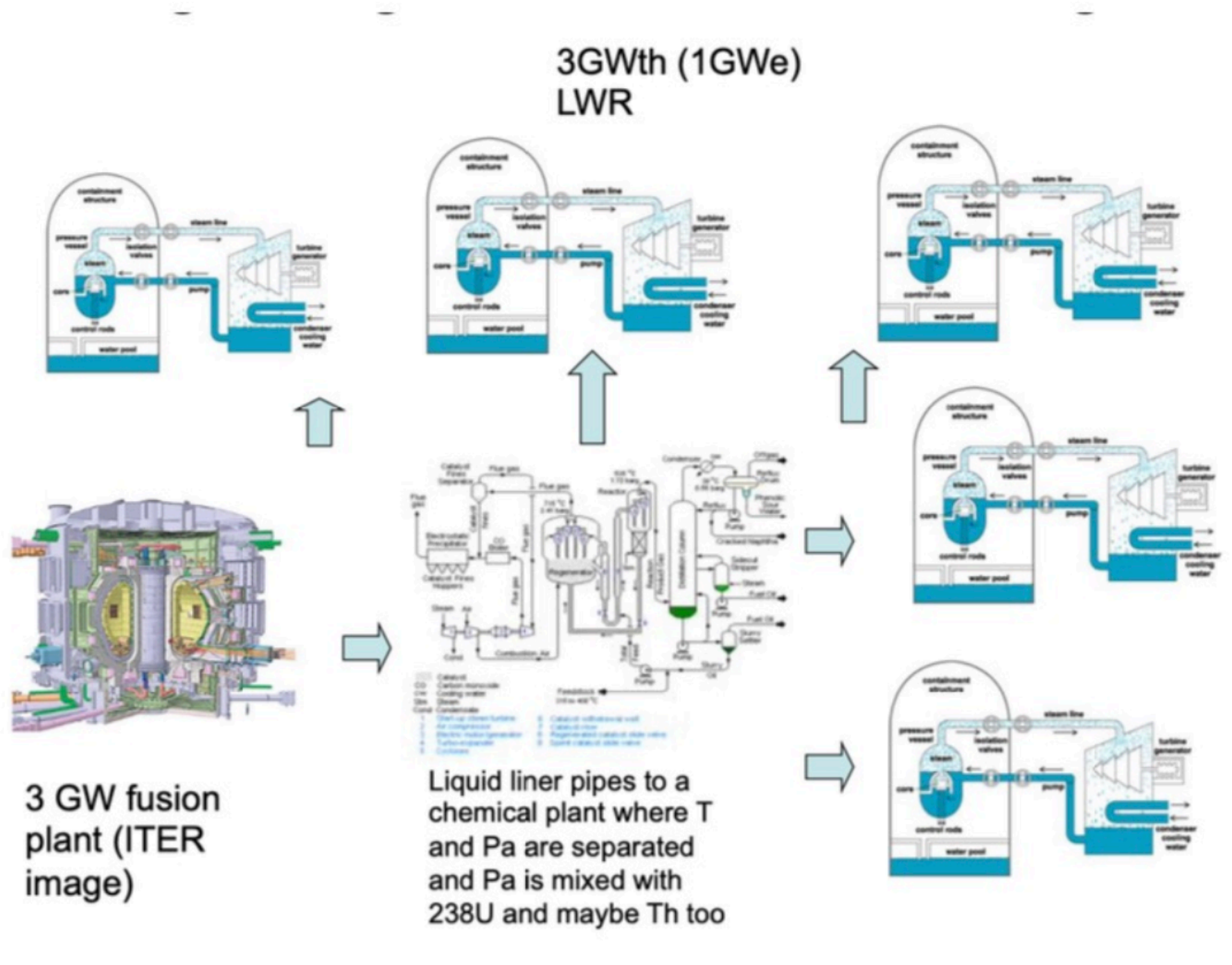

Figure 5: One fusion breeder (a tokamak in the figure) can fuel at least 5, and possibly as many as 10 thermal reactors. This is a powerful argument that fusion should be considered not only for energy, but also for breeding, which, as we will see, make many fewer demands on whatever the fusion system is.

Starting from scratch, it is impossible to know at this point whether fast neutron breeders, thermal thorium breeders, or fusion breeders fueling thermal reactors, or some combination of all three is the optimum for future power for the world. However, one thing is for sure. Only fusion breeders can rescue thousands of 'stranded' thermal reactors which are 'out of gas', just like electric vehicles on a cold winter day in Chicago (Finley). Even a children's book (Bonnell), (with slight modification) sees this, as illustrated in Fig (6).

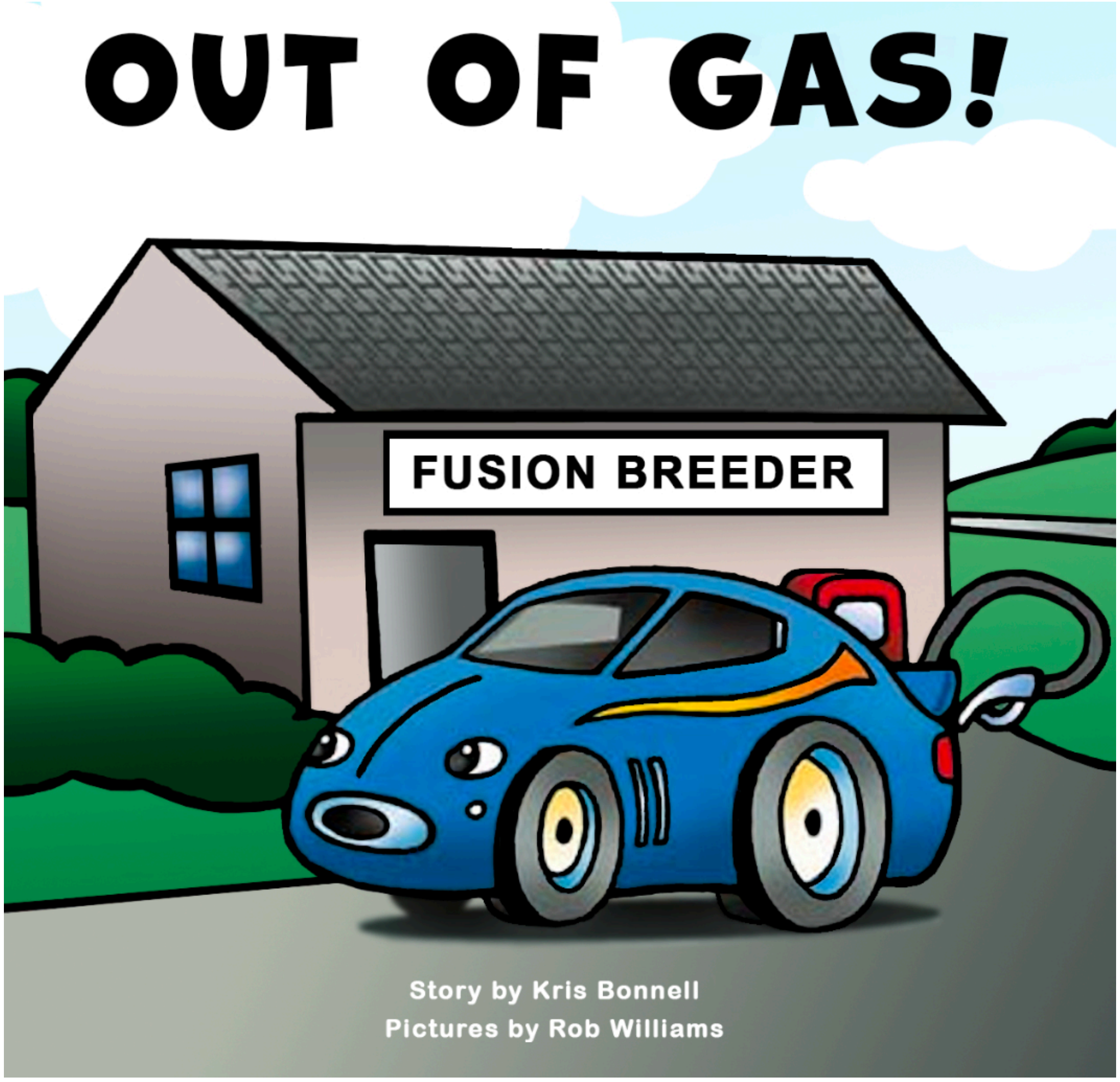

Figure 6: Of all breeding options, only fusion breeding can 'refuel' thermal reactors once they 'run out of gas' and are stranded.

## Section III: Fusion research within the lines

The fusion project is a very large scientific project, run by a very large bureaucracy, the Department of Energy in the United States, and other like departments around the world. In the US, the bureaucracy is divided into two branches, one which supports magnetic fusion for energy, and another which supports laser fusion for nuclear weapons simulation.

### III. A: Fusion reactions:

To orient ourselves, we now list the most common possible fusion reactions. The most important of these is the DT reaction, it has the highest reaction rate and requires the minimum plasma temperature.

$$D+T\rightarrow\rightarrow n(14.1MeV) + He(3.5MeV)$$

The deuterium for this reaction can easily be supplied by the world's oceans and the amount is basically unlimited. About 1/6000 hydrogen atoms is deuterium. A similar reaction uses helium 3 instead of tritium, but because of the additional Coulomb repulsion, requires higher plasma temperature and has a lower reaction rate.

$$D+{}^{3}He\rightarrow\rightarrow p(14.7MeV)+He(3.6MeV)$$

One problem with of each of these reactions is that neither tritium nor helium 3 (in usable quantities) exists on earth. Tritium must be bred, and helium 3 exists on the surface of the moon. The fusion project is currently considering only breeding tritium. Tritium can be bred from lithium, and there are two possible breeding reactions. The first is exothermic:

$$n+{}^{6}Li\rightarrow\rightarrow T(2.75MeV)+He(2.05MeV)$$

The second possible reaction is endothermic, taking 2.47 MeV away from the reacting particles:

$$n+{}^{7}Li\rightarrow\rightarrow T+He+n\ (-2.47\ MeV)$$

Clearly this reaction requires an energetic neutron. However, depending on the breeding blanket and the end use, it may be worth the energy price to price to preserve the extra neutron.

It is also important to note that tritium is not itself a stable nucleus. It is unstable to a decay into helium three, and its half-life of 12 years, or ~8% is lost each year. Hence if the blanket absorbs the neutron and forms the tritium, one would like to extract it as quickly as possible. It would be much better if the tritium could be extracted immediately as it is produced. For this reason, a flowing liquid blanket would appear to have a great advantage over a solid blanket. The liquid would flow from inside the fusion reactor to a chemical separation plant, which would remove the tritium, and then insert it back into the reactor.

Also note that the basic nuclear reaction produces only a single neutron. If there are absolutely no losses, this is all that would be needed to resupply the tritium for the tokamak. However, there are always some losses. If the blanket is a solid and it is replaced yearly, that alone will give rise to a loss of 4% of the tritium. The saving grace here is the large energy of the fusion neutron. There are many reactions, called spallation reactions, where a high energy proton or neutron generates more neutrons. We will discuss some of these later. Even for pure fusion, it is essential to rely somewhat on spallation for neutron multiplication just to recover any losses. For fusion breeding, it is absolutely essential, as the single fusion neutron needs more neutrons to recover losses and to breed both a triton and a fissile nucleus, say $^{233}U$.

A reaction not requiring any breeding is the DD reaction, which may proceed along one of two paths with equal probability for each.

D+D→→n(2.5MeV)+3He(0.8MeV) or

D+D→→p(3MeV)+T(1MeV)

This reaction produces less energy and requires still higher plasma temperature. One might look at it not as a reaction to produce energy, but to breed tritium and helium 3. However, to do this, one must find a way to remove the $^{3}He$ and tritium nearly instantaneously from the DD plasma, as the DT reaction rate from the generated tritium is at least 100 times greater than the DD rate for all plasma temperatures below 1 megavolt. In Figure 7 are shown reaction rates for the three fusion reactions.

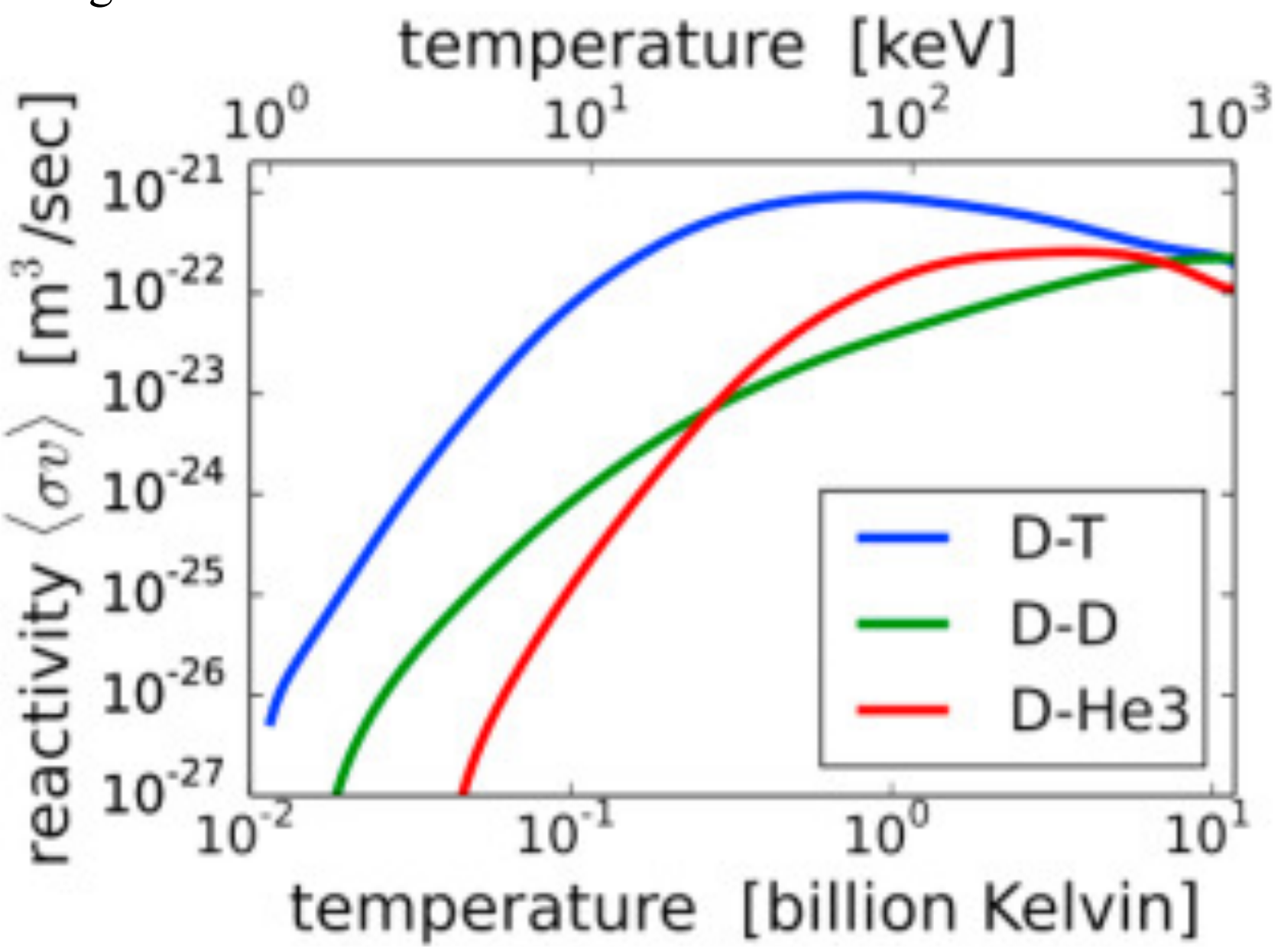

Figure 7. Fusion rate for the three fusion reactions.

Clearly the DT reaction rate is largest and requires the lowest plasma temperature to proceed. The reaction rate maximizes at a temperature about 50 keV. However, the total reaction per unit volume goes as $n^2\langle\sigma v\rangle$. If the pressure is constrained to some certain value, i.e the density is this pressure divided by the temperature, then the reactions maximize at the temperature where $\langle\sigma v\rangle/T^2$ maximizes, or at about 16–17 keV, where the reaction rate is about $\langle\sigma v\rangle \sim 3 \times 10^{-22}$ $m^3/s$.

III. B tokamaks:

The main effort of the American magnetic fusion effort through the years has been the tokamak. This is a toroidal shaped containment vessel containing a hot, current carrying plasma. This plasma is confined by both the toroidal magnetic field set up by large superconducting coils external to the plasma, as well as the magnetic field generated by the plasma current itself. The toroidal current is driven inductively by setting up a toroidal electric field driven by an increasing vertical magnetic field in the hole in the center of the torus. That is, the plasma forms the secondary coil of what is a simple transformer. If course one can only increase this vertical field so much, or in other words, the transformer is powered by only so many Volt-seconds, after which the plasma ends. This plasma is heated by the Ohmic heating by the transformer, as well as by neutral beams and/or microwave and/or millimeter waves injected from the outside.

It is immediately obvious that tokamaks have a problem going to steady state operation, which is certainly necessary for a reactor. There have been extensive studies on externally driving the current with neutral beams, microwaves, and millimeter waves; theoretical studies in the US, and on actual tokamaks in China and Korea. However, they also have another problem, disruptions. This is the sudden termination of the discharge, with the plasma energy smashing into the wall. The parameter space where the tokamak can operate disruption free is pretty much known. Within this parameter space, a tokamak, JT-60 has operated disruption free for at least 30 seconds (Ishida, Isayamab), although how much longer they can do this has not been experimentally verified at this point.

In the tokamak world, there is a simple parameter which is generally accepted to be a figure of merit of the tokamak. This is called the triple product; that is the product of the number density in $m^{-3}$ times the confinement time in seconds, times the plasma temperature in kiloelectron Volts (keV). It is roughly proportional to the fusion power divided by the drive power, usually defined as the Q of the reactor. Over the 30 years from 1970 to about 2000, this parameter showed a steady increase as more and more tokamaks were built, tokamaks getting larger and larger, and tokamaks people were learning more and more about. In about 2000, the triple product maximized at the JT-60 tokamak in Japan at a value of $1.6 \times 10^{21}$.

At the time the tokamak community took justifiable pride in the fact that its triple product increased in time at about the same rate as the number of circuits on a computer chip. This is illustrated in Figure (8). Notice that the Princeton Plasma Physics Lab (PPPL) had many entries, (ST, PLT, PDX. and TFTR), as did the Japanese with various versions of their JT tokamaks.

There have been were 3 large tokamaks constructed.  These were TFTR in Princeton (Hawryluk), JET in England (Gibson), and JT-60 in Japan (Kishimoto); large meaning they ran with up to 40 MW of external power (neutral beams or microwaves) and had a major radius more than 2.5 meters.  They each got reasonable fusion power, typically with a Q of about 0.25 for various modes of tokamak operation.  The Japanese JT-60 tokamaks, about 90 cubic meters of plasma, never ran with a DT plasma, owing to the difficulties of dealing with tritium, but in DD plasmas, got a Q of 1.25 had it been a DT, instead of a DD plasma.

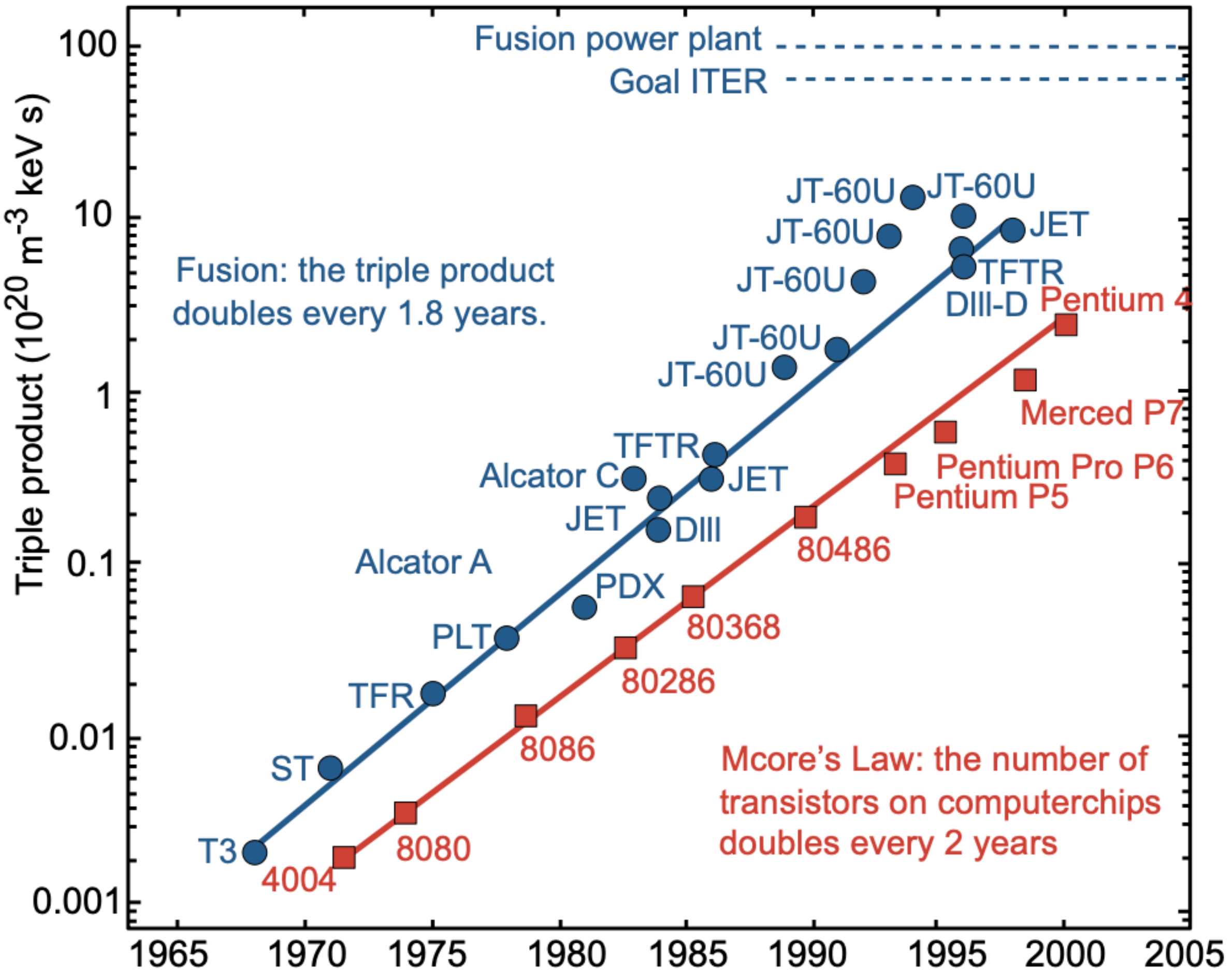

Figure  8:  The increase in triple fusion product as a function of year, as opposed to the number of transistors on a computer chip.  However, at each stage of the red curve, industry was able to produce a competitive product.  In 2000, the blue curve leveled off, needing larger tokamaks to advance further, tokamaks which were unaffordable for a single nation or group of nations.

Since the next level of tokamaks was unaffordable, the major nations decided to build one as a cooperative venture among the world's more advanced nations.  This concept was originally discussed between President Ronald Reagan, and Soviet Communist party head, Mikhail Gorbachev at one of their summit meetings.  This evolved to ITER (International tokamak

Experimental Reactor) to be built by the United States, the Soviet Union, the European Union, China, Japan, and South Korea. It was estimated to cost ~$10B and be a machine with an 8-meter major radius, a 2.7 meter minor radius, a poloidal elliptical configuration with a vertical elongation of ~ 1.7, a magnetic field of 5 Tesla (T), and volume of about 2000 cubic meters. The hope was that with 150 Megawatts of external power, the DT plasma would give a Q of 10 and produce 1.5 billion Watts (1.5GW) of fusion power for 400 seconds (Aymer). We call this tokamak Large ITER. The United States pulled out of the collaboration thinking it was too expensive.

However, the remaining partners decided to reduce the size of the machine in the hope of enticing the United States to get back in. The US rejoined, and later, India came in as a full partner. The new design had a 6-meter major radius machine with a volume of 1000 cubic meters (Campbell, iter). The hope is to produce a machine which will still give a Q of 10, but now with 50 Megawatts of input power producing 500 MW of fusion power, also for 400 seconds. It will be extremely interesting to see if ITER can run at full power disruption free for this time. The capital cost was now estimated at $5B, half the original estimate.

Finally in 2002, the partners agreed to build it and agreed on the preliminary design. However, the question then became where to put it. Both Japan and the European Union put in strong proposals. The United States, Japan and Korea voted for Japan; the European Union, China and Russia voted for Europe. The voting was a tie for 3 years (India was not then a full voting partner). Then India joined as a full member and broke the tie in 2005, by voting for Europe. The original capital cost of ITER was to be $5B, with first plasma in 2016, and fusion by 2025. However, the delays and cost overruns have been unmerciful. The capital cost is now estimated as $25B with first plasma in 2025, and fusion experiments to be complete by 2040. While many of the partners may now wish they originally had second thoughts, the construction of ITER is now ~ 70% complete. Most likely there is no going back now. We can all hope that ITER meets its milestones. Figure (9) is a schematic of the ITER tokamak and a photograph of the construction site.

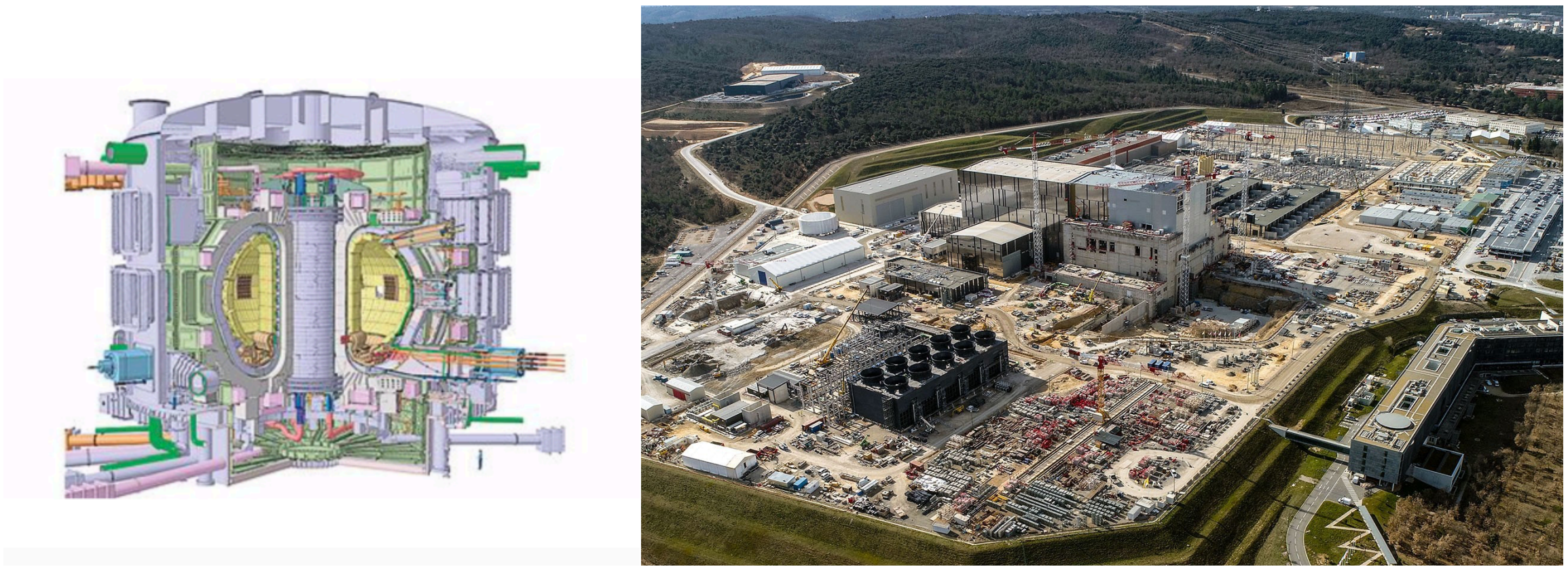

Figure (9): An artist's conception of the ITER tokamak. The distance from the center of the torus to the center of the plasma is 6 meters. Also shown is a photo of the ITER construction site when the construction was ~70% complete.

III C:  Stellarators:

Stellarators were one of the first fusion devices investigated.  The stellarator does not rely on a plasma current but attempts to produce properly shaped magnetic surfaces externally with various configurations of external wires.  The Princeton Plasma Physics Lab (PPPL) originally in the 1960's, concentrated on helical windings of around the outside, but the confinement was terrible.  When the Russians, in the late 1960's showed that a tokamak had much better confinement, the PPPL almost immediately switched to tokamaks.  The main issue is that the tokamak is a two-dimensional confinement, while the stellarator is inherently a much more complex three-dimensional configuration.  That is the tokamak had azimuthal symmetry around the torus, while the stellarator did not.

However, labs in Germany and Japan never gave up.   The main stellarator effort now is at the Max Planck Garching Lab in Germany (ipp, Boozer A, Fleschnir,  Boozer B)). Using theoretical work in the United States which showed how to make this three-dimensional configuration as 'nearly two dimensional as possible', they were able to greatly improve the confinement.  Also, unlike the tokamak, the confinement depended only on external magnetic fields which one can control; it did not rely on the often-uncooperative plasma to help confine itself.  Neither is there any need for a transformer, with limited Volt-seconds to drive any current, so developing a steady state stellarator is generally thought to be easier than developing a steady state tokamak. Also, stellarators are thought to have less of a problem with disruptions than tokamaks. However, the price paid was that the configuration was much more complicated, and it was a much larger configuration for a much smaller plasma volume, and it had considerably worse confinement.  At this point in time, the Germans seem to have the most modern stellarator, the Wendelstein-7.  A schematic of the coil configuration and the device itself is shown in Figure 10 (ipp).

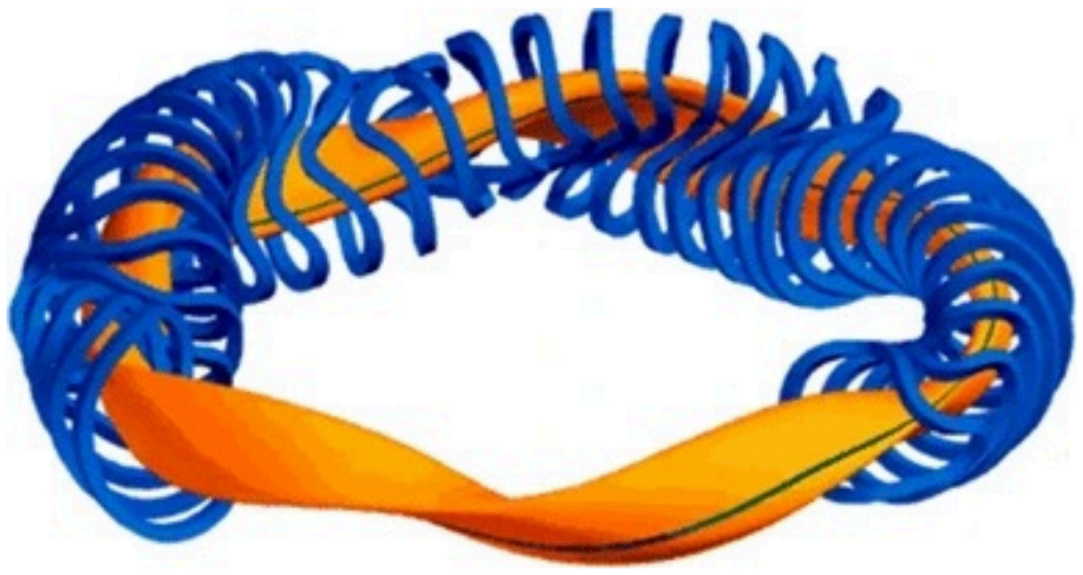

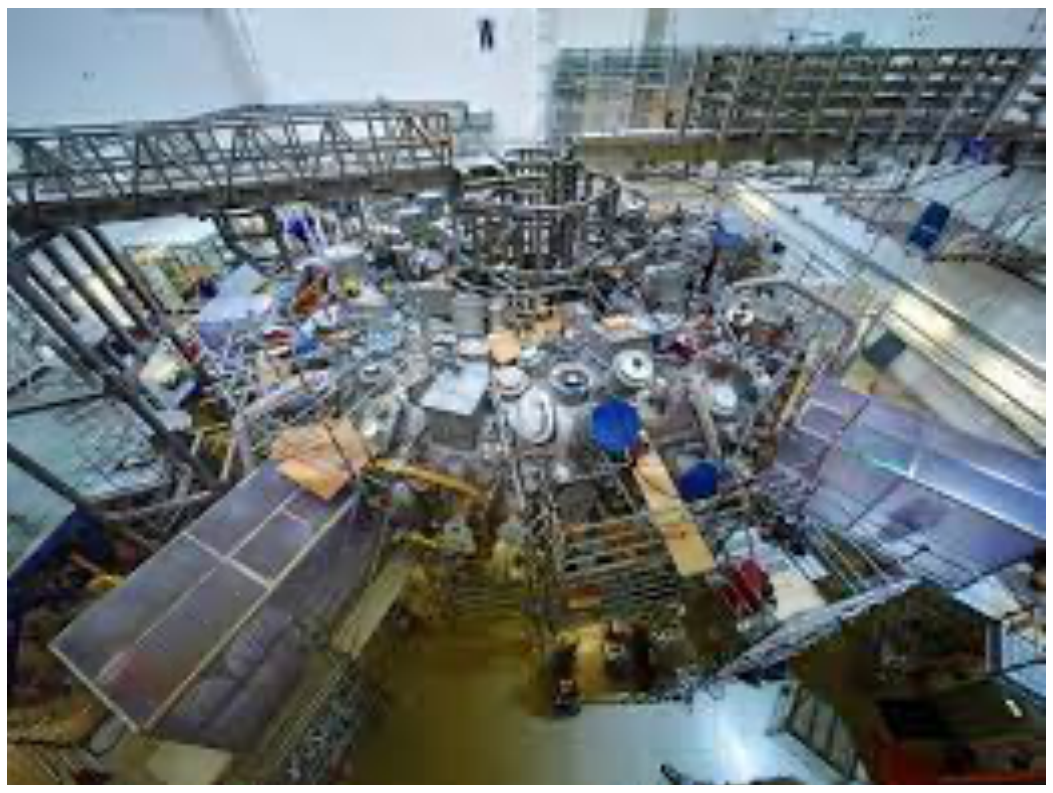

Figure (10)  Left:  The Wendelstein-7 Stellarator's coil configuration.  In blue are shown some of the 70 superconducting coils, and in orange, the basic plasma shape as one goes around the torus.  It has a 5.5 meter major radius, a half meter minor radius, and has a plasma volume of 3 cubic meters.  The superconducting coils are 3.5 meters tall.    Right:  A photo of the stellarator in the lab.

To compare the confinement, recall that JT-60, with a plasma volume of 90 cubic meters has a triple product of $1.6\times10^{21}$; while Wendelstein-7, with a 30 cubic meter volume achieved $6\times10^{19}$, about a factor of 30 less. However, Wendelstein-7 does have a larger triple product than does any confinement configuration except a tokamak.

## III D: Laser fusion

Magnetic fusion has the goal of energy for the world and is managed by the energy part of the US Department of Energy. Laser fusion currently has the goal of becoming a laboratory for the study of nuclear weapons and is managed by a completely different part of the US DoE.

Almost as soon as lasers were invented, scientists thought of them as drivers for inertial fusion. Initially the thought was to simply deposit the energy in a target, heat it to fusion temperature and let it fuse. However, the laser energy needed was enormous, many, many megajoules. A significant theoretical breakthrough came when Nuckolls showed that by ablatively compressing the target, the laser energy could be enormously reduced, perhaps to as low as 10 kJ or less (Nuckolls). Ablative compression means is that the laser deposits its energy in the outer region of the target, which heats up, ablates away, and the inverse rocket force compresses the remainder of the target to fusion conditions. This compresses the inside of the target to a tiny dense hot spot so that fusion can begin. The neutrons escape, but the alpha particles are absorbed locally and heat the surrounding region, so that they begin to fuse. In other words, the laser initiates an alpha driven burn wave. It plays a role more like a spark plug, which only initiates the fuel ignition. Notice that the alphas play a vital role in laser fusion.

To achieve this requires a spherical implosion, so maintaining the spherical symmetry is of utmost importance. This means that one must find a way to minimize the effect of the Raleigh Taylor instability, which is unavoidable, since ablative compression necessarily means the acceleration of a heavy fluid by a light one. An enormous effort has been made here, and the community generally agrees that the outward ablative flow has a strong stabilizing effect, although just how strong is still under study. In any case, by taking advantage of the flexibility one has in designing the target and the laser pulse, one can exert a measure of control over the flow to minimize the effect of the instability.

While Nuckoll's idea is still the main one being pursued today, as we will see, his original estimate of necessary laser energy was nothing if not optimistic. In the pursuit of laser fusion, LLNL embarked on major program developing a series of larger and larger lasers, Argus, Shiva, Nova, Beamlet, and finally the National Ignition Facility (NIF), a Megajoule laser. All of these are Nd glass lasers with a wavelength of 1.06 μm. However, at such long wavelengths, laser plasma instabilities become a major worry. Accordingly, LLNL has developed frequency multiplication techniques to operate at third harmonic, about 1/3 μm wavelength. LLNL now routinely operates with pulses more than a megajoule at third harmonic. The University of Rochester Laboratory for Laser energetics (URLLE) has also taken this approach with their OMEGA laser (30 kJ). The US Naval Research Laboratory (NRL) has taken a different approach, using an excimer KrF laser at a wavelength of 0.248 μm with its NIKE (3–5 kJ) and Electra lasers. More recently, on a shoestring budget, it has developed a 200 Joule ArF excimer laser having a far ultraviolet wavelength of 0.193 μm (Obenschain).

In terms of economics and timelines, the experience of NIF has not been so different from the experience of ITER, but on a much smaller scale. It was approved in 1995, to be finished in 2002 at a cost of $1.1B. Instead, it was finished in 2009 at a cost of $3.5B. Once NIF became operational, the 3-year National Ignition Campaign (NIC) began. Its goal was to achieve Q=10, by 2012.

The NIC configuration specified the target placed in large spherical target chamber with a diameter of about 10 meters, and with many entry ports to give access to the laser beam and various diagnostics. The laser target is at the center of this chamber, which is illustrated in Figure (11) from the LLNL web site.

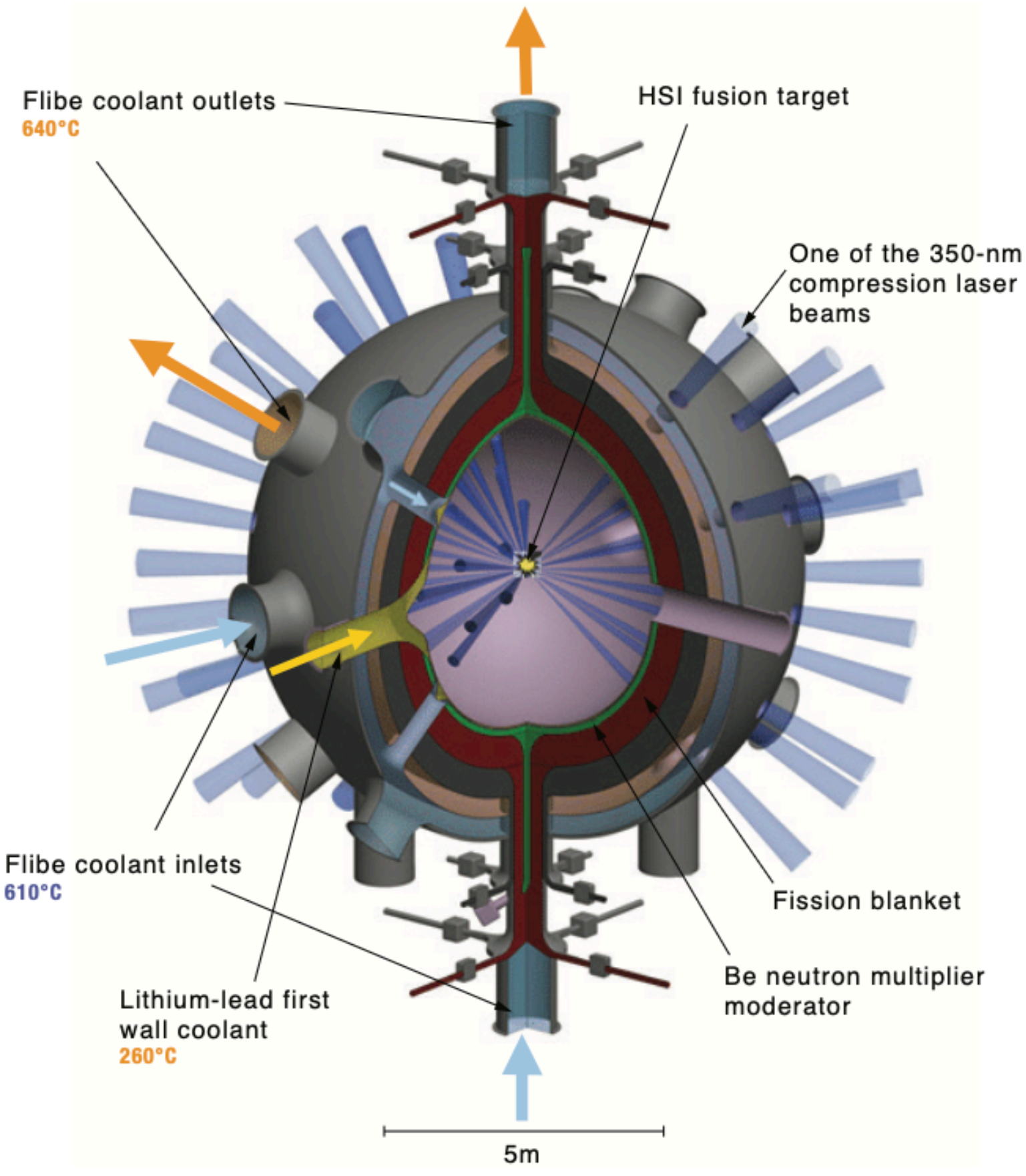

Figure 11: The 10 meter diameter target chamber for the NIC campaign.

The National Ignition Campaign uses what is called indirect drive. That is the target is placed inside of a container called a hohlraum. The laser illuminated the inner high Z walls to produce a black body of temperature 250-300 eV, producing an intense X-ray flux, which irradiates and implodes the target. A schematic of their configuration is shown in Figure 12, (Zylstra B)

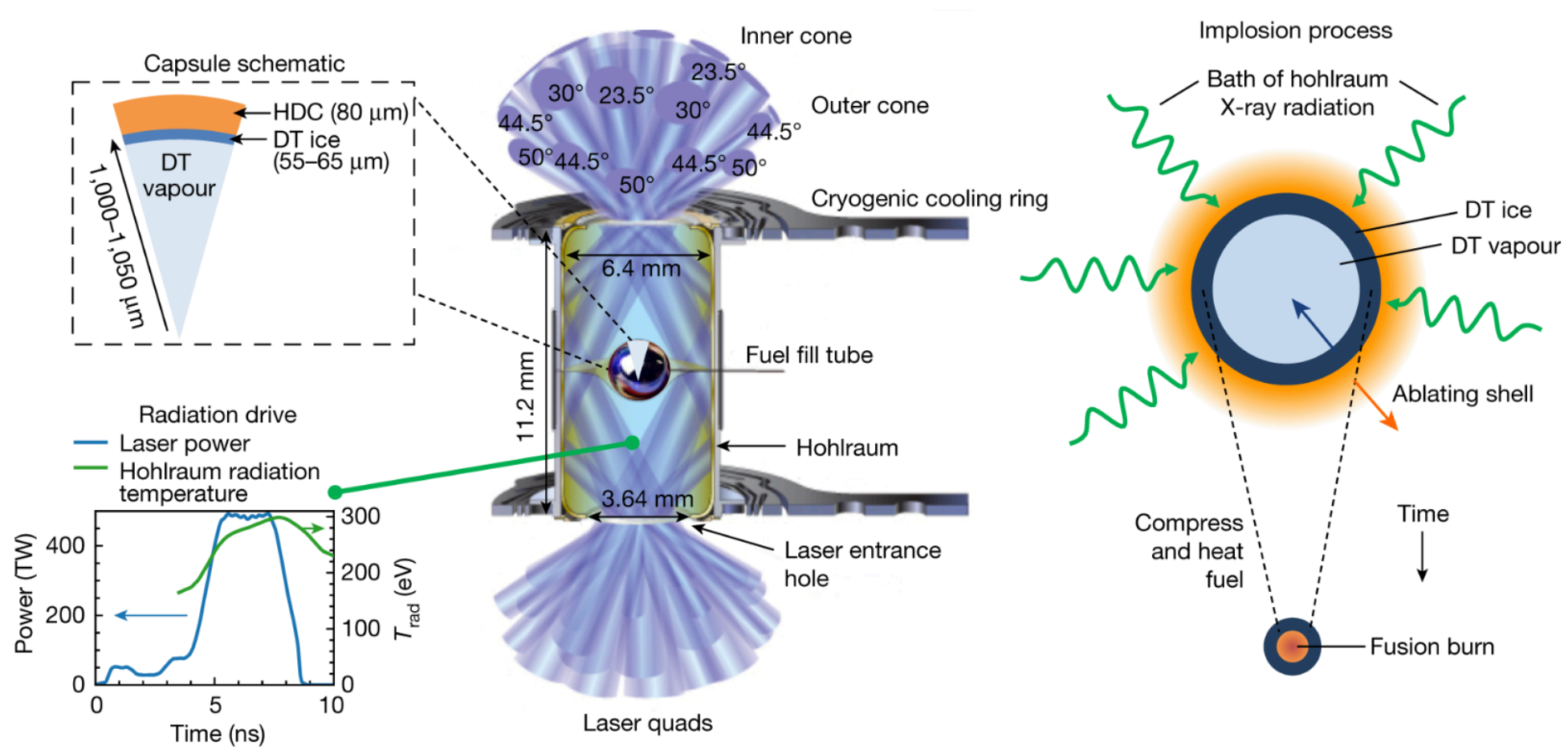

Figure 12: The configuration of the LLNL NIF successful experiment on producing an ignited plasma

It is important to note that the laser is off for the final phase of the target implosion. That is the laser gets the implosion up to maximum velocity, and then turns off, so for the final phase of the implosion, the target is coasting in. The whole idea of laser fusion is for this implosion, at what is called the 'bang time' (when the laser is off), to create a small thermonuclear burn in the center of the implosion. The 14 MeV neutrons escape, but the 3.5 MeV alphas are absorbed locally and heat the surrounding plasma. This allows the possibility of a burn wave, or in other words an ignited plasma.

However, when the laser was turned on in 2009, the target did not cooperate. Early experiments showed gains of only ~1%, three orders of magnitude less than predicted. The LLNL group studied and studied their system and gradually got the Q up to ~10%. At this point, in 2021, lightning suddenly seemed to strike. To their surprise (many of their diagnostics were set for lower levels and saturated) and delight, they suddenly got a gain of ~70%. While this is still considerably less than their original goal of a Q of 10, they got an unmistakable signature that they had produced a burn wave, 9 years after they had hoped to do so.

Look at their measurement of fusion production as a function of time, from 3 shots, after the time of maximum compression. It is reproduced in Figure 13 taken from (Zylstra A).

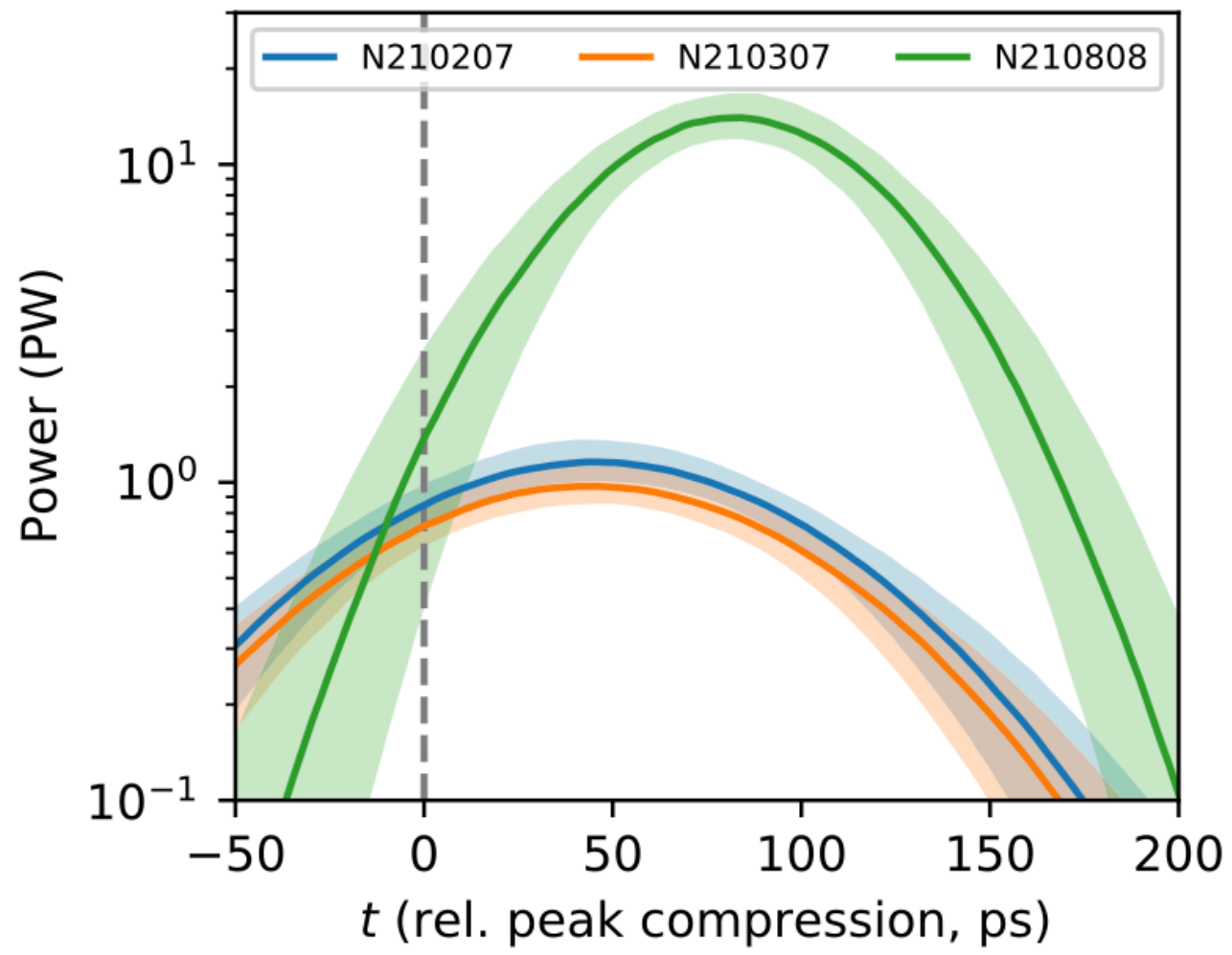

Figure 13: Fusion power vs time, relative to peak compression (t=0, gray dashed line). The shaded bands denote 1σ uncertainty.

Notice that the maximum fusion power is *after* the bang, in other words the peak fusion is from an *expanding* plasma. This could only be from setting up an alpha heated burn wave.

However, nature did not give up her secrets easily. After this successful shot, they tried to repeat it, and for a year they failed. Early in that year, they arranged to have a plenary talk at APS-Division of Plasma Physics (DPP) meeting for the fall of 2022 (Divol), a year after their first real success. It was to be given by one of their scientists, Laurent Divol. But for most of the year, there was nothing to talk about. When Divol wrote his abstract, which was published in the bulletin, it was all about the reason for failure. But then suddenly in October, about a month before the meeting, when it was too late to alter the abstract, they succeeded again, this time getting a gain of ~ 60%. Reading the abstract, and listening to the talk, one would think they were about two very different experiments. Then a few months later they had an even better success, getting a Q of 1.5, with 2 MJ laser light producing 3 MJ of fusion energy.

Further evidence of an alpha burn wave can be gleaned from their direct measurements of radius and temperature of the expanding plasma. These have been presented in an online seminar (Seminar) and in Laurent Divol's conference plenary talk (Divol), but apparently have not been yet written up in the archival scientific literature. I had been at both online presentations, and my sketches of their results are shown in Figure 14 below.

Shot Aug, 2021, Laser energy 1.7 MJ, fusion energy 1.3MJ, Q=0.76! On line seminar, Sept 2021

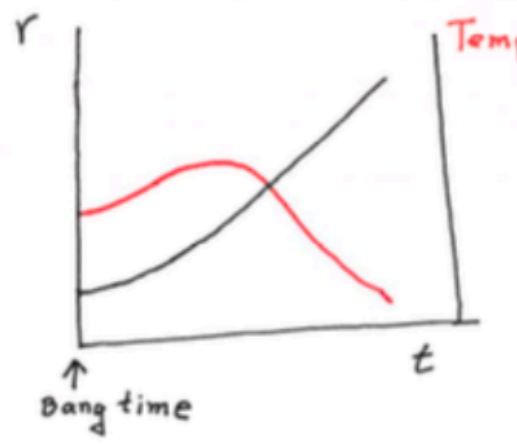

Shot Oct, 2022, Laser energy 1.9 MJ, fusion energy 1.2 MJ, Q=0.63! Laurent Divol, plenary talk DPP meeting.

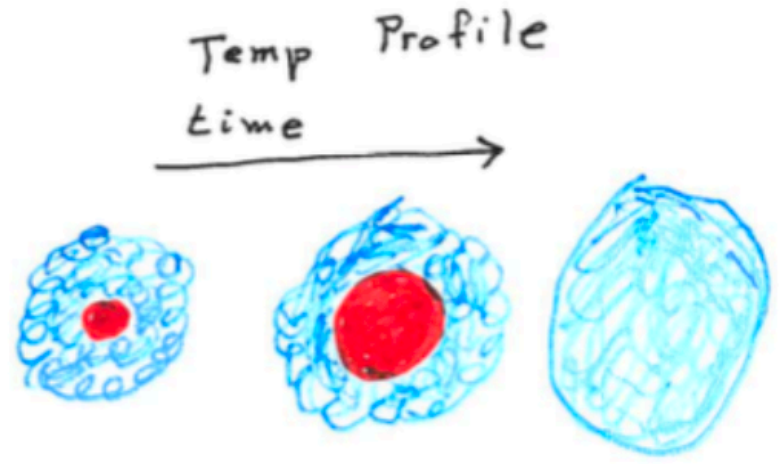

Figure 14: Other LLNL diagnostics of the alpha burn wave.

Apparently, so far, they have had 3 such successful shots, the first with a Q = 0.76, the second with a Q =0.63, and the third with a Q =1.5, producing 3 MJ of fusion energy.

I believe the (LLNL) result of getting a Q =1.5, and more importantly, demonstrating an alpha burn wave, in a laser fusion target in an indirect drive configuration is a breakthrough for the ages. They achieved this, at a much lower cost, nearly 20 years before ITER hopes to do anything like this. I believe that 100 years from now, it will be regarded as one of the most important experiments of the 21st century, and this author is pleased to congratulate them on this remarkable achievement. The Secretary of Energy was at the presentation of their results. At the high holiday service at my synagogue, the rabbi, when presenting some of the good news of 5783, mentioned this result!

Furthermore, I believe that there is a reasonable, but obviously a difficult path from the LLNL success to the fusion energy for the world. However, I believe that this path is a different one, and with a different goal, than the one the Lawrence Livermore National Lab is embarked on. Hence, I make the case that it is appropriate, no, not only appropriate, but required that DoE set up a different lab to investigate this intriguing possibility.

Section IV: Some problems with fusion within the lines

IV A: Tokamaks

There are quite a few obstacles a tokamak would have to get over before becoming an economical power plant. These are:

1. Let's say that ITER is successful and produces 500 MW of fusion power with 50 MW of Ohmic, beam, microwave, and millimeter wave power. A standard nuclear reactor converts thermal to electrical power with an efficiency of ~ 1/3, so ITER would deliver ~170MW to the grid. However, beams and microwaves are also produced with typically 1/3 efficiency as well, meaning that the ITER would draw ~ 150 MW from the grid, leaving nearly nothing for anyone else. The ITER web site recognizes this problem and mentions that ITER itself is only a first step to the DEMO, a follow-on reactor that will generate economic power. The DEMO must be smaller, cheaper, more powerful, ~ 3GWth, and have a higher Q, at least ~40. It is doubtful that now anyone has a credible design of the DEMO, and realistically this is not even possible without knowing the results from ITER. Who knows how long will take to develop the DEMO, if indeed if this is possible at all.

2. It is still uncertain (to say the least) that a reactor grade tokamak knows how to generate the current. The transformer can only drive it for some finite interval. There have been tokamaks which have been driven for long periods of time with microwaves and neutral beams in China (Experimental Advanced Superconducting Tokamak [EAST]) (Wan, Xiang) and Korea (Korea Superconducting Tokamak Advance Reactor [KSTAR]) (Jung-Gu). However, these had very poor confinement, or looking at the other way, it took too much power to drive the currents. EAST and KSTAR have achieved triple products of $\sim 10^{19}$, about where Ohmic driven tokamaks were 40 years ago. To put this in perhaps more understandable terms, JT-60 confines ~ 10 MJ of plasma energy, powered by 10-30MW of external power. EAST and KSTAR contain ~ 200 kJ of plasma energy, powered by ~ 5-10 MW of external power. It may be that these tokamaks will find more efficient ways of generating the current externally, but at least at present, the result are rather discouraging.

3. Tokamaks are constrained by what this author has called 'Conservative Design Rules', or CDR's. This paper will not go into detail about what they are, but they are carefully spelled out in the references. They are limits on the density, pressure, and current the tokamak plasma can contain. The penalty for violating these rules is usually a disruption, which of course is not tolerable in a reactor. Each individual element of the CDRs, for instance the limit on plasma pressure, is well known. However, taken together, the conclusion is both astounding and ignored. As far as this author is aware, he was the first, and nearly the only one to point out these implications (Manheimer 2009, 2104, 2018, 2020 A, 2021 A, 2023 A). The author is aware of only one other paper making similar points (Freidberg 2015). If these constraints persist, and no way is found to get around them, as the 50-year experience with tokamaks would indicate, then the size of the tokamak will almost certainly be too large to be economical. A tokamak generating 3 GW of thermal fusion power, like a conventional coal or nuclear reactor, with a 5T field would need a major radius of at least ~ 10 meters. In other words, taking its coils into account, putting one end of the machine at the goal line of an American football field, the other end would be at about the 30-yard line. If the field is 9T, which might be possible with the modern high temperature superconducting tapes, the major radius would be at least ~5 meters.

The parameter limiting the plasma pressure is called $b_N$, or the normalized beta. It is proportional to the plasma pressure divided by the current. The constant of proportionality depends on the toroidal magnetic field and the geometry. The conservative design rule is that it must be less than about 2.5. Once this limit is exceeded, the plasma is unstable to what are called ballooning modes. The parameter limiting the plasma current is called the safety factor, or $q_{95}$, which is proportional to the reciprocal of the current. The constant of proportionality depends on the toroidal magnetic field and the geometry. The conservative design rule is that $q_{95}$ must be greater than about 3. Once $q_{95}$ gets less than 3, the plasma becomes unstable to what are called tearing modes. There is a great deal of tokamak data confirming these limits, especially on the two largest tokamaks, JT-60, and JET. A sample of this data (Kishimoto, Ishida, de Vries) is shown in Fig (15).

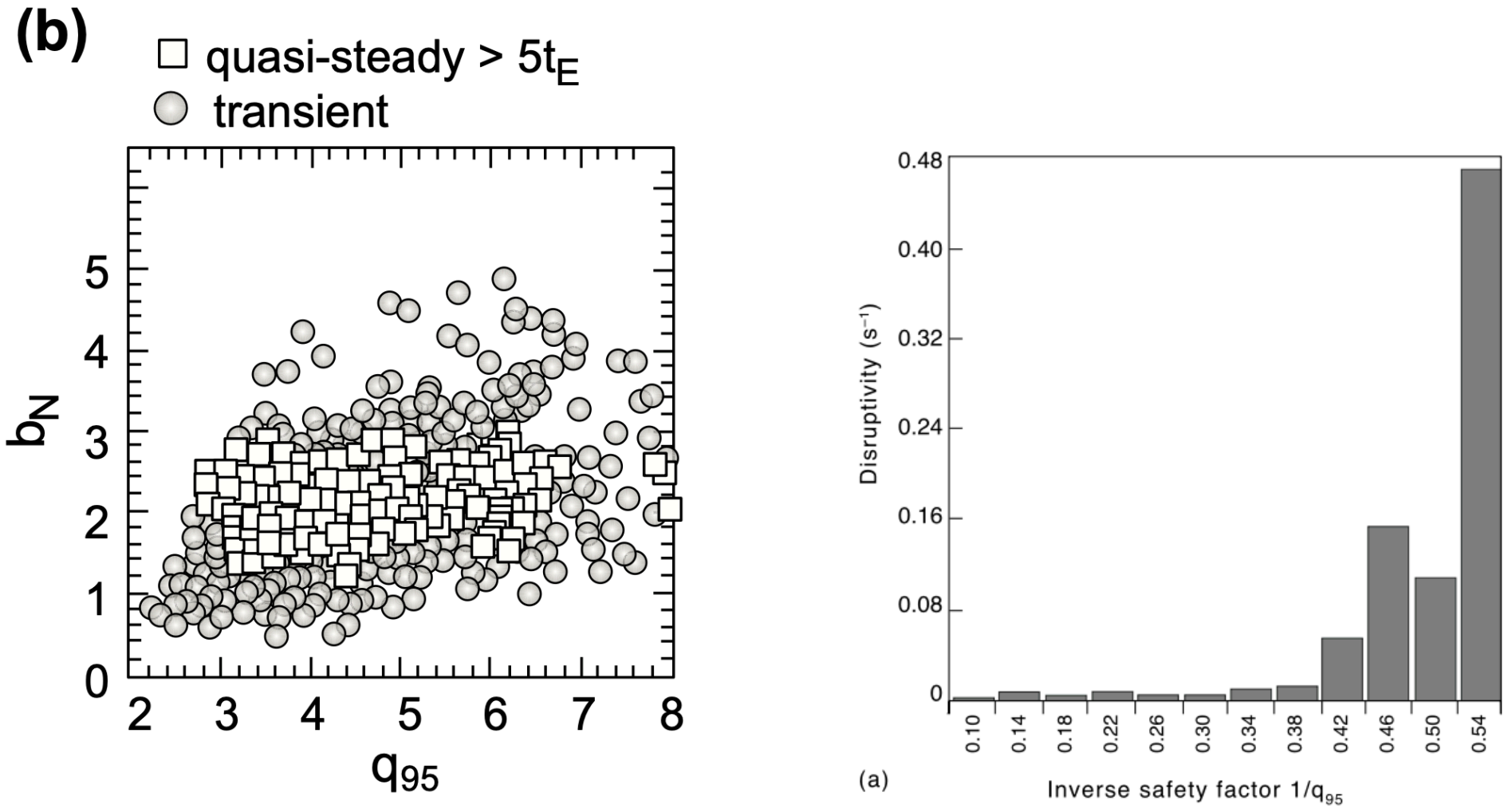

Figure 15: Left hand side is data from JT-60. It is a map in ($b_N$,$q_{95}$) parameter space of the characteristics of many discharges. The hollow squares represent discharges that were steady state for as long as the for as long as the discharge lasts, a time greater than 5 energy confinement times. The solid circles are for discharges that abruptly ended in some way. The most fusion relevant steady state discharges are around $b_N$ ~2.5 and $q_{95}$ ~3. This figure clearly verifies the CDR's for many JT-60 discharges. Higher $q_{95}$ steady state discharges are lower current discharges and do not have the density and temperature relevant for fusion. The right-hand side is data from JET. From analyzing many discharges, they examined the time between major disruptions as a function of the reciprocal of $q_{95}$. The reciprocal of this time, in $sec^{-1}$ is plotted as a function of $1/q_{95}$. Clearly once $1/q_{95}$ gets larger than 0.38, the disruption rate very quickly increases, again confirming CDR's. While disruptions may not appear to be a problem for smaller values, a disruption rate of 0.01means a disruption about every 2 minutes, clearly not a viable rate for a reactor. It will be very important to see if ITER can routinely operate at full fusion power, disruption free for the full 400 seconds (~7 minutes).

4. What does one do with the alpha's produced? Unlike the 14 MeV neutrons which escape and hit the wall, the 3.5 MeV alpha will stay in the plasma. What effect will they have? Do they benevolently heat the plasma at just the right amount to maintain the proper temperature, no less and no more? How can one control this heating? Ultimately, the alphas must be removed. How does one do this without also removing the D and T? Very possibly, in fact, it seems the most obvious possibility, they will simply remain in the plasma, building up its pressure, i.e. the $b_N$, until it gets above 2.5, and causes a major disruption. While there certainly must be many paper studies, there is no experimental data guide us. Basically, the tokamak program seems to regard the alphas as a nuisance; at this point, they are.

5. The first wall is struck by 14 MeV neutrons at the very least, but also by radiation, and possibly with fast neutrals and ions. Who knows what will come back into the plasma? Recycling of wall impurities into the plasma could be a very big problem. Since there has not been a fusion reactor yet, it is not clear how big a problem this is, but given the impurities from the wall in non-reacting plasmas, one can certainly expect it to be a major issue in reacting plasmas.

6. ITER's plasma, which we do not understand very well, has the energy of a 100-pound bomb. Who knows what effect a major disruption would have on the device? Its magnetic field, produced by superconducting coils, has the stored energy of a one-ton bomb. While uncontrolled quenches are very rare, and there are many controls, they do happen. Once the CERN accelerator suffered an uncontrolled quench, and the accelerator was down for a year until the machine could be repaired. If a disruption, or anything else, would cause an uncontrolled quench in the confined space of ITER, it would probably take down the building and much more. While this is certainly not a probable occurrence, one must keep in mind the fact that ITER is storing an enormous amount of plasma and magnetic energy in a rather small volume. It is a real potential safety concern.

7. ITER will require a tremendous amount of tritium to do its experimental work, tritium that is expensive and not readily available. Some tritium is produced naturally, but this supply is much too dilute to obtain from, say sea water. The relative concentration of tritium to hydrogen is estimated as $10^{-18}$. Most tritium used is produced by CANDU heavy water reactors. These use deuterated water as the moderator instead of ordinary water as in an LWR. A loss mechanism in the LWR is the absorption of the neutron to produce deuterium. The analogous loss mechanism in the CANDU is the absorption of neutron by the deuterium to produce tritium. Each CANDU reactor produces ~ 130 grams, about $3 \times 10^{25}$ tritons, per year, or about $10^{18}$ tritium nuclei per second. There are ~40 CANDU's worldwide producing about $4 \times 10^{19}$ tritons per second. A single 3 GWth fusion reactor would need to burn ~$10^{21}$ tritium atoms per second, so there is no way that CANDU reactors can supply a fusion economy, or even a single fusion reactor. Once a fusion economy becomes established, it can easily enough produce the tritium needed. However, in the experimental phase, there is no fusion economy, and supplying the tritium needed for the 5-year experimental phase of ITER could put a serious strain on the tritium supply. The 1000 $m^3$ of ITER, at a tritium number density equal to the deuterium number density would need ~$5 \times 10^{22}$ tritium atoms (about a quarter of a gram) just to fill up the machine for a for a single experiment. Recovering unburned tritium after each shot is quite important. The JET project has done some work on this (Peacock). However, as other large experimental fusion reactors come online, and as ITER gains experience, so there is less and less unburned tritium, there will be less and less available tritium. The supply from CANDU reactors may or may not be sufficient. *As a sideline effort, the fusion research project, whatever it is or may become, should vigorously oppose every effort to take any CANDU reactors offline. Very soon, these will be playing a major part in the fusion research effort. Without them, it is difficult to see how fusion could ever progress.*

8. The experience of ITER is that it has hit one delay and cost overrun after another. The capital cost is now at least 5 times the initial estimate, and who knows what additional cost overruns and delays the project will encounter (Kramer A). While this author feels that there is no choice but to complete the project, it is hardly an encouraging harbinger.

IV B. Stellarators

Stellarators start out with the fact that at least at this point, they do not demonstrate the plasma confinement that tokamaks have. However experimental data so far implies that long lived discharges in non-reacting plasma are not especially difficult to achieve, although the more energetic the stellarator, the less experience with long lived discharges.

Many of the points made about tokamaks above apply to stellarators as well, and most likely to any steady state magnetic fusion configuration also. Points numbers 1, 4, 5, 6 and 7 certainly apply to stellarators. As there is less experience with stellarators, it is not clear whether point 8 applies or does not, most likely it will. Point 2 does not apply to stellarators, as they do not carry a current. The main question is point 3. Surely there are pressure and density limits on stellarators, but they have not been publicized in a general, easily found form. Conservative design rules were not publicized until 2009 (Manheimer 2009), but the author had sufficient experience with the tokamak project to figure them out, and could sort through the experimental data to see that they were valid. These had major impacts on the likelihood of tokamaks becoming economical reactors. I do not have the required experience or knowledge to do the same with stellarators. Very likely the theory and experimental results are not sufficiently mature for anyone to know what these limits are, but surely there are pressure and density limits. This author certainly encourages experts on stellarators to come up with these limits, whatever they are; limits not only in theoretically derived, but limits which are backed up by considerable experimental data, as they are for of tokamaks.

While we are unable to consider these limits, we can do some simple scaling. We know that Wendelstein has a major radius of 5.5 meters, a plasma volume of 30 cubic meters, and the magnetic field is produced by 70 superconducting coils, each one 3.5 meters tall. Assuming the reacting stellarator plasma is like that of a tokamak, we would expect that the plasma volume would have to be ~4000 cubic meters for a 3 GWth reactor, or about 130 times the volume of Wendelstein. The simplest way to achieve this is to multiply each linear dimension by about a factor of 5. Hence, we are talking about a 28-meter major radius, and 70 superconducting coils each one about 18 meters tall, about the height of a 4 story building. Using the football analogy again, if one end of the stellarator is on the goal line of an American football field, the other end would be on about the 20–30-yard line of the *opposing* team. Each field coil would reach up to about the second tier of the stadium. *This does not sound cheap!* While more compact stellarators designs might become possible, and perhaps paper studies indicate this, the experimental data at this point indicates that a 3GWth stellarator reactor will be much too large to be economical.

IV C The privately funded 'fusion start-ups'

The private sector has jumped into the fusion quest. Many companies say that they will produce commercial electricity via fusion in the next decade or so. This author is on record as saying that they will all fail (Manheimer 2023 A) and, so do references cited in the Introduction. The obstacles between where the project is now, and what is needed, are just too great, really a mile high and a mile wide. If fusion is ultimately successful, private companies will obviously play a big role in its development. However, this author feels quite sure that its involvement is vastly premature at this point. Many other have made the same case.

This paper will not consider these further, but only lists many quotes of these companies, quotes that promise fusion after such and such time, a time which has already long since passed, and there is still no fusion.

From Geek Wire, Oct 23, 2023:

"Almost a decade ago, Helion predicted reaching scientific breakeven by 2017."

"Zap hoped to get there this (scientific breakeven) year (2023), though it almost certainly won't."

From Jassby, FPS April 2019:

"Tri Alpha says it will produce a working commercial reactor between 2015 and 2020," (Tri Alpha, now called simply TAE, started in ~ 1998, so in 25 years it has produced no commercial fusion.)

"GF targets prototype by 2015 and a working reactor by 2020"

"Lockheed will have a small fusion reactor prototype (power plant) in five years…and a commercial application within a decade," (claimed in 2014)

IV D: Laser fusion:

Since this paper is an argument for a switch from magnet fusion to laser fusion, we first consider the difficulties of tokamak fusion which any laser fusion configuration does NOT have (numbers correspond to Section IV A)

4. Unlike magnetic fusion, which does not know what to do with the alphas, and generally regards them as a nuisance, laser fusion knows exactly what role the alphas play. They are an integral part of the concept, they set up an alpha generated burn wave, and the LLNL NIF experiments have already demonstrated this.

5. Unlike magnetic fusion, laser fusion has no problem with recycling. The fusion burn occurs in less than a nanosecond. Any fusion product which dislodges wall material, by the time this wall material gets to the burning plasma, the fusion reaction has long since been complete.

6. Compared to magnetic fusion, laser fusion stores very little energy. Consider NIF for instance, which combines the radiation of 192 independent lasers. Let's say that each element stores a 2 MJ, equivalent to half a pound of TNT and one of them blows up. It will not necessarily cause more than a small amount of local damage. A laser necessary for economical fusion will necessarily be much more efficient than NIF, so the energy storage will be much less. Laser fusion is inherently much safer than magnetic fusion.

7. The laser targets as illustrated in Figure (12) contain about $10^{19}$ tritium atoms. This is about one five thousandth of the tritium it takes to fill up ITER. Hence with the tritium necessary to fill ITER for a single series of shots, NIF (or any other laser configuration) can do about 5000 shots. Research on laser fusion will put much less strain on tritium supply.

8. The LLNL NIF development has also suffered significant cost overruns and delays, but nothing like what ITER has put up with. NIF was approved in1995, to be finished in 2002 at a cost of $1.1B. Instead, it was completed in 2009 at a cost of $3.5B, about triple the initial estimate. ITER was approved in 2005 at an initial cost of $5B, with construction to be completed in 2016. Its cost is now estimated as $25B, 5 times the initial estimate, and the expected completion date is 2025.

However, despite their recent successes, the LLNL approach has many obstacles to overcome if the object is energy rather than nuclear security and simulation of nuclear weapons. As the goal of the project is nuclear stockpile stewardship, and not energy, the sponsor is only interested in X-ray drive. It is not interested in things like laser efficiency, rep rate, average power capability, and bandwidth; parameters important for energy but not so much for nuclear simulation. It is certainly neither interested in direct drive, nor capability to track and focus on a fast-moving target.

Furthermore, each shot involves a hohlraum, a precisely engineered container, made with expensive materials like gold or uranium and currently costing thousands of dollars each. While mass manufacturing of hohlraums will undoubtedly bring down their price considerably, even if the target produces a total energy of ~100MJ, which would translate to ~33MJ of electric energy, or ~ 10 kWhrs, worth about a dollar, it gives a very low-price limit for the ultimate economically acceptable hohlraum price. Second, only a small fraction of the laser light (in the form of X-rays) makes it to the target; the rest is lost through other channels. This is shown in Fig (16) taken from the LLNL publication (Hybrid).

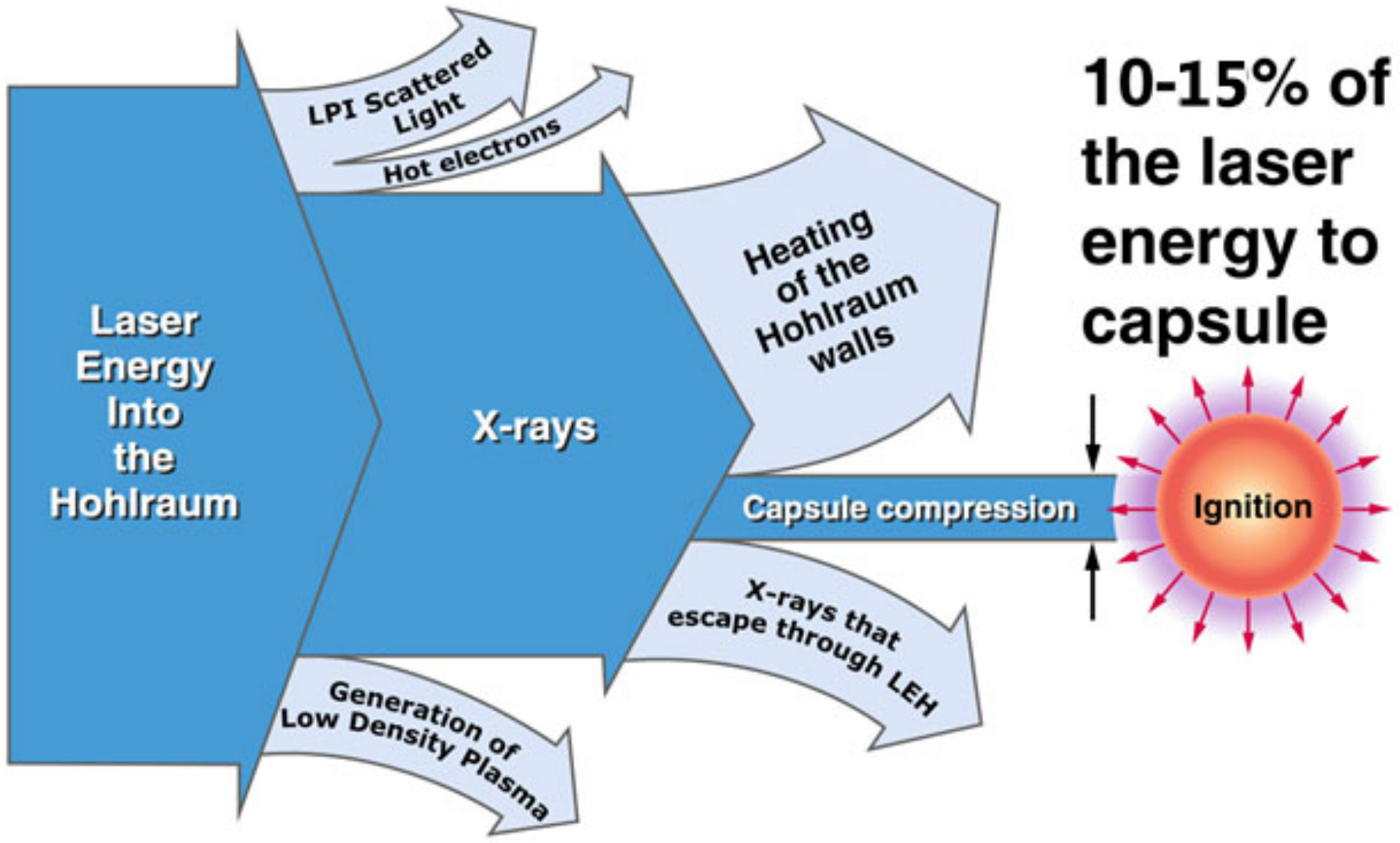

Figure (16): A schematic of where the laser energy goes for an indirect drive configuration. Only 10-15% of the laser energy makes it to the target in the form of X-rays.

Finally, the LLNL configuration is fine for one shot, with the target on a small stalk or in a transparent ‘tent’. Focusing the laser on it is relatively simple. It is rather like hitting a golf ball on a tee. To do this continually, targets would have to be continuously shot in a high speed, with each shot certainly traveling in on a slightly different path. The target engagement becomes more like hitting a variety of Jacob DeGrom’s fastballs, curve balls, sliders, changeups….., *on every pitch*. Not only does the target have to be in the right place, it must have the proper orientation also, so the laser is aligned with the axis of the hohlraum, or to use the baseball analogy a bit further, the batter has to hit the ball at a precise phase of the ball’s spin. Laser fusion is playing baseball, not golf!

Section V: Laser fusion, coloring outside the lines

The first element of laser fusion outside the lines, is an alternative approach which is called direct drive, where the laser light directly hits the target, and without any of the losses shown in Fig (17). This is a schematic of direct drive configuration taken from (nrl). Since the target is a sphere, it does not have to have any specific orientation, so the target engagement becomes much simpler.

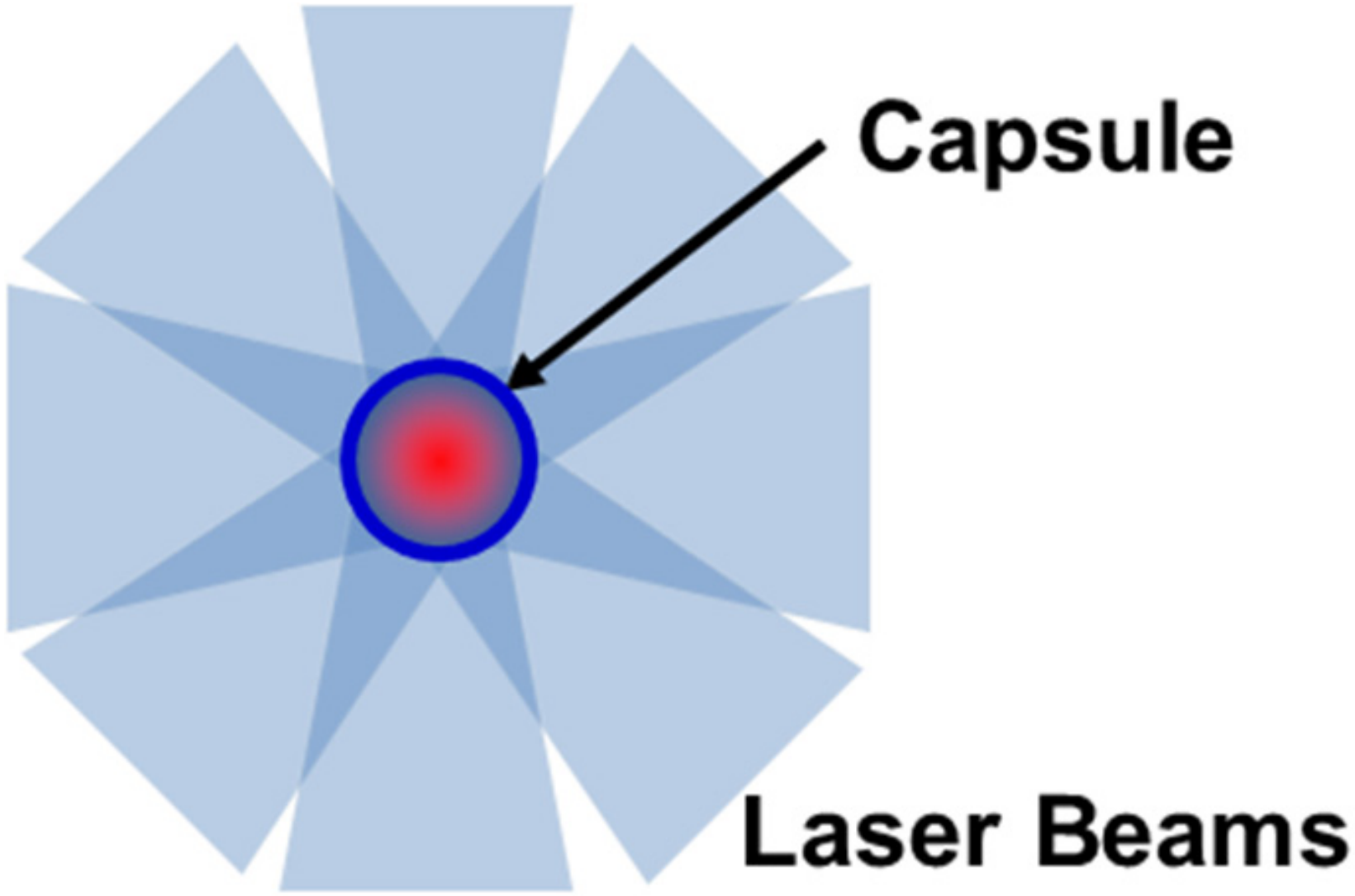

Figure (17): A schematic of direct drive laser fusion taken from (nrl). The laser beams directly hit the target, so very little energy is wasted in other loss channels, as is the case with indirect drive.

It is possible to perform experiments with nearly 4π illumination on cryogenic DT targets. The University of Rochester Laboratory for Laser Energetic (URLLE) has done direct drive experiments with its smaller OMEGA (Ω) laser, with cryogenic DT targets [Sangster]. While with their 30 kJ laser they could not achieve ignition they did get decent neutron production and central ion heating. Figure (18), taken from (Sangster) shows the neutron production and central ion temperature as a function of implosion velocity. The maximum neutron production is about $2 \times 10^{13}$, or about 45 J. Since the maximum energy of OMEGA is 30 kJ, this corresponds to a Q of at least $1.5 \times 10^{-3}$, not that much less that what NIF has achieved with on its initial indirect drive experiments.

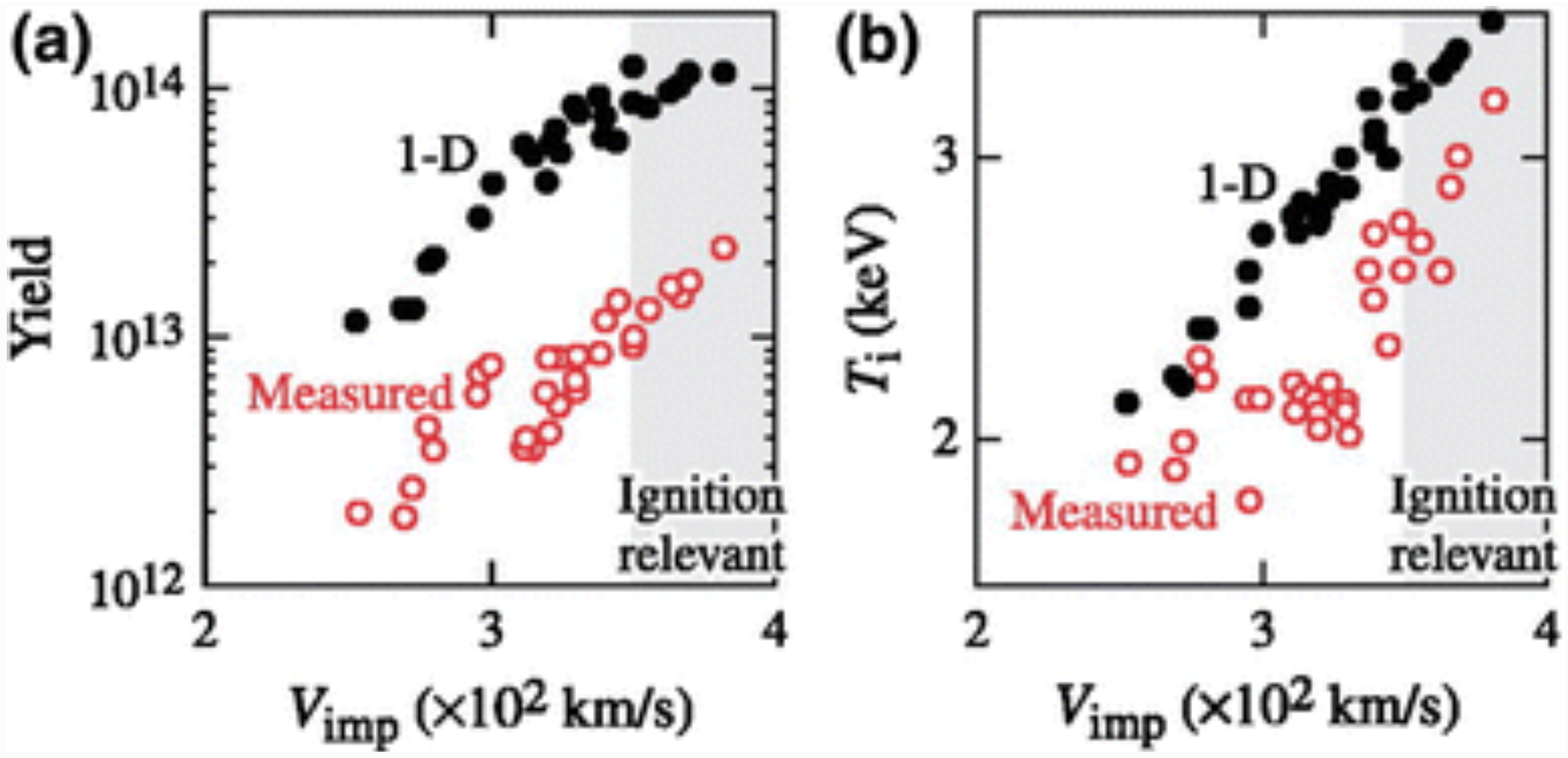

Figure (18): Measurements of neutron yield and ion temperature in direct drive powered by the OMEGA 30 kJ laser at URLLE. (Sangster)

However, both LLNL and URLLE have done the work with frequency tripled Nd lasers having a wavelength of 333 nm. The Naval Research Lab (NRL) has taken a different approach; for decades, they have for decades used excimer lasers. Their first one was NIKE, a KrF laser, producing ~2-3 kJ of ultraviolet light with a wavelength of 248 nm. Figure 19 is a photo of the spherical target chamber for the NIKE laser.

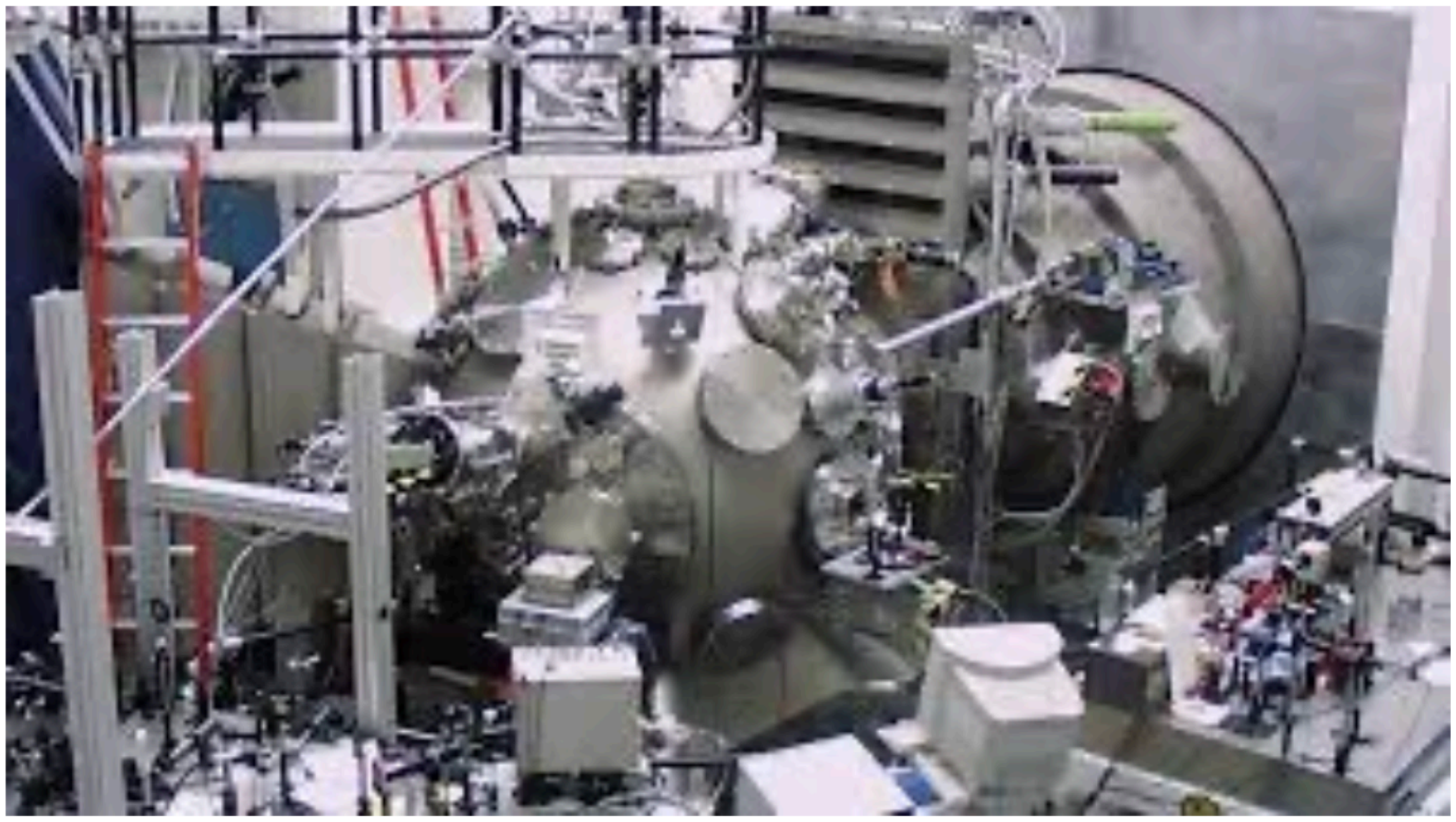

Figure 19 The spherical target chamber (~ 1 meter in diameter) for the NIKE laser at NRL

Also they have developed a rep rated KrF version called ELECTRA. This produced a laser pulse of 300-700 Joules at a rep rate of 2-5 Hz. The goal of ELECTRA was 'large enough to be convincing, but small enough to be manageable'. It believed it achieved this goal. It has generated over a hundred thousand shots at various rep rates and for various run times.

ELECTRA was an important element of the High Average Power Laser (HAPL) program, a multi-institution, multi task project managed by NRL.  It existed from 1999 until 2008, when it became a casualty of the financial collapse. Its accomplishments have been documented on the ARIES web site (ucsd) and in a published journal article (Sethian). The program goal was to investigate every aspect of laser fusion with the goal of developing all the science and engineering necessary to make it a reality. For instance, its namesake goal was to develop lasers capable of high average power and efficiency, which are suitable as drivers for laser fusion. To bring this about, HAPL supported 2 laser projects, the Electra laser at NRL, and Mercury laser at LLNL. The latter is a frequency tripled diode pumped solid state laser, with an energy of 50 J per pulse and runs at about 10 Hz.

However, lasers were hardly the only aspect investigated in HAPL. It involved some 30 institutions investigating such things as first wall, final optics, the chamber, the target manufacture, the target positioning, the target tracking…. The program was an integrated program with the goal of achieving economical and practical laser fusion. No single goal was regarded as paramount. In fact, the mantra of the project was "You cannot solve your problem if you make the next guy's problem impossible".  The HAPL program had made steady progress during its brief lifetime, along a very broad front, and believed that there were no showstoppers.

More recently NRL has converted ELECTRA to an ArF laser, producing a laser beam of ~ 200 J with a far ultraviolet wavelength of 193 nm (Obenschain).  Figure (20) is an photo of ELECTRA as an ArF laser.

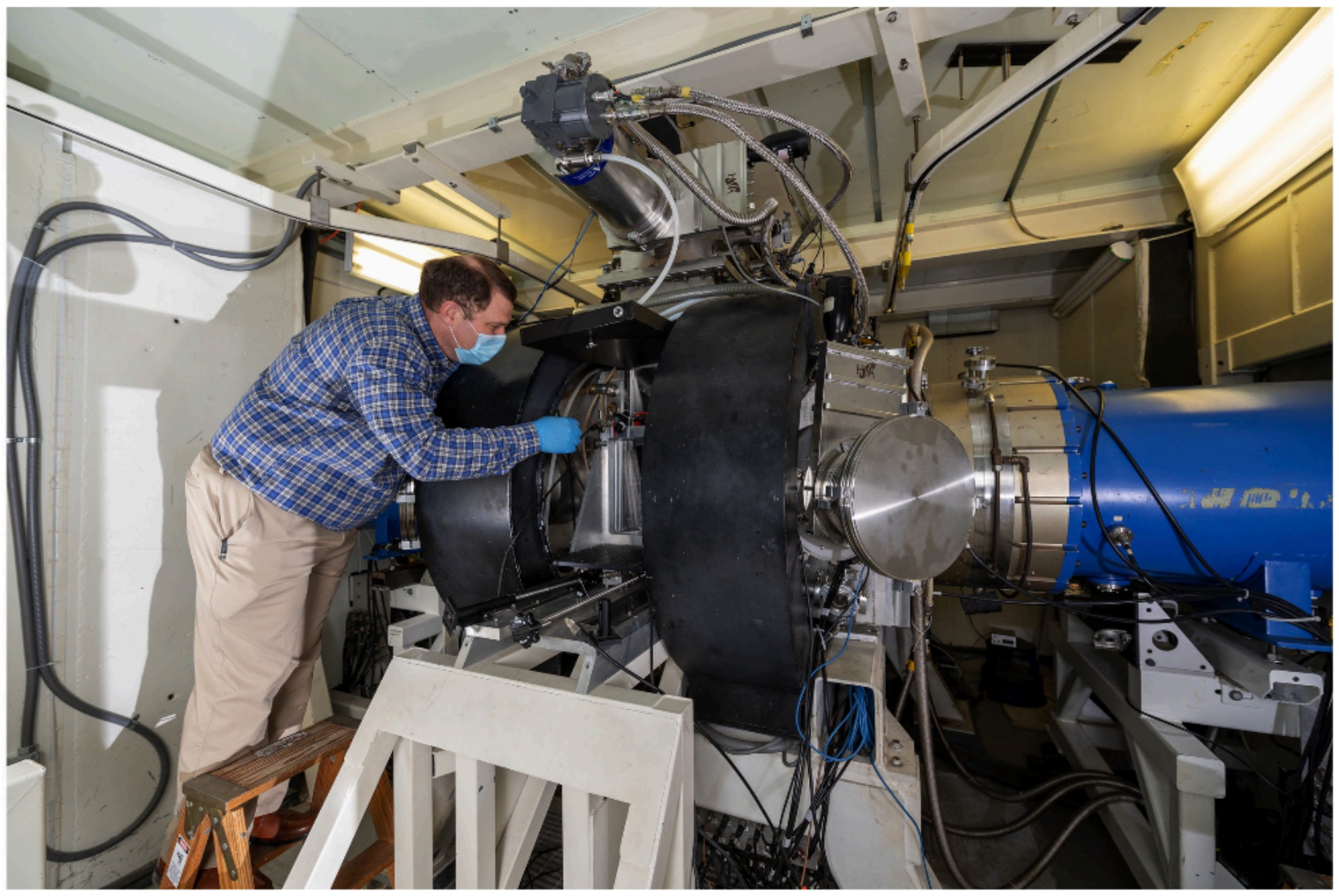

Figure 20   A photo of the ELECTRA ArF laser at NRL.  The laser produces a reprated ultraviolet pulse of ~ 200 J at a wavelength of 193nm.  At this point it is the  ArF laser relevant for laser fusion,  with the largest average power.

NRL has also calculated the gain of laser fusion targets as a function of laser energy for 3 different laser sustems, 333nm, 248 nm and 193 nm; and for two differerent laser pulse types. The results of these calculation are shown in Fig (21).

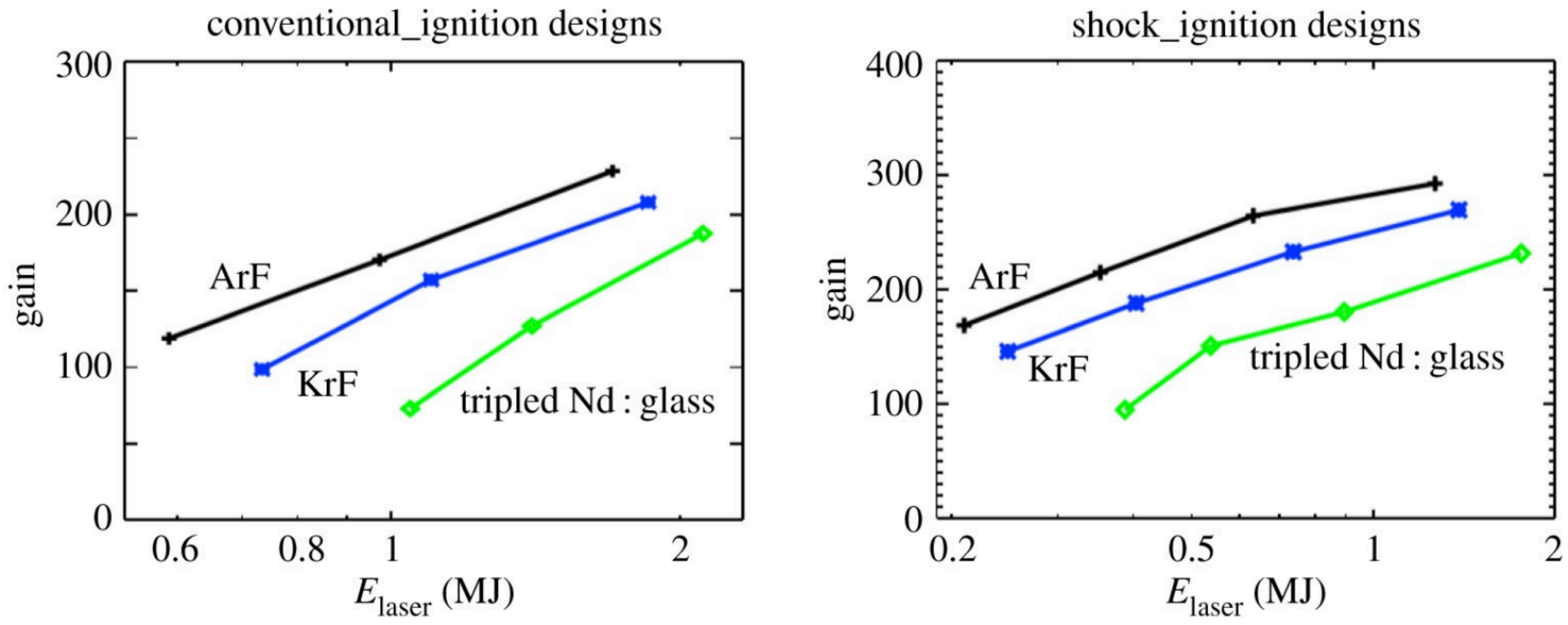

Figure 21 NRL calculations of gain as a function of laser energy for 3 different laser wavelengths for two different laser pulse shapes. Excimer lasers' calculated gains are larger than that of frequency tripled Nd lasers, in part because the shorter wavelength light reaches further into the target and produces produces a higher ablation pressure.

The excimer lasers' shorter wavelength produce larger gain, generating larger ablation pressure at higher plasma density. However this is hardly their only advantage over Nd glass lasers. Excimer lasers use a flowing gas as the lasing material instead of a solid as is the case of Nd lasers. Hence there is much less danger of optical damage to the amplifying material. This has been a near constant head ache for LLNL as they pushed up the laser energy. Not only does the use of a flowing gas most likely make it easier to have higher energy pulses, it also most likely makes it easier to have a high laser rep rate and average power. Furthermore, excimer lasers have demonstrated both higher efficiency and higher bandwidth. (Obenschain) claims that ArF lasers ought to have wall plug efficiencies of ~ 10% and bandwidth of ~ 10 Terahertz (THz). This bandwidth is likely required for stabilizing a variety of laser plasma instabilities. Finally, excimers have the capability of zooming. This means they can vary their focusing properties in a controlled way during a pulse. Hence as the target implodes, the laser can focus on the shrinking size of the target.

As stated in the last section, LLNL's goal is nuclear stockpile stewardship, and not energy. That sponsor therefore is only interested in X-ray drive and is not interested in things like efficiency, rep rate, average power capability, bandwidth, or ability to track a fast moving, wobbling target, parameters important for energy but not so much for nuclear simulation. Because the demands on the laser for nuclear simulation and energy are so different, it makes sense that the nuclear simulation and energy goals be handled differently. As stated in the Introduction, because of the success of LLNL in demonstrating an alpha burn wave, but with the goals of the nuclear simulation and energy so different, it makes sense to set up another DoE lab dedicated entirely to direct drive fusion, most likely with an excimer laser. This lab would both cooperate and compete with LLNL. However, because the goals of the two labs are so different, the competition would be minimized, and the cooperation would be maximized. At this point, I believe it is also fair to say that NRL realizes that it will not be that lab. The Navy is simply not sufficiently that interested in power for the civilian sector that it can devote the necessary resources necessary to such a large project. Also, it is even less likely that DoE would put support of the necessary magnitude in a non-DoE lab. However, I feel relatively sure that NRL would be more than ready to cooperate in setting up this new lab.

For now, let us take NRL's gain calculations at face value. Namely we will consider a 2 MJ laser with a gain of 250, produced by a laser with a 10% efficiency. If this is pulsed 6 times per second, it would produce a fusion power of 3 GWth, just like a conventional coal or nuclear plant. With a conventional generator, this would produce electricity with an efficiency or 1/3, or produce power for the grid of 1GWe. However, each pulse of the 10% efficient laser would take 20 MJ, or an average power of 120 MW. Hence the circulating power, 120 MW, is a relatively small amount of the grid power produced by the laser, it seems like a viable system. In the next section we will see that this picture may not be so rosy.

## VI. Fusion Breeding, coloring far outside the lines:

As mentioned in the Introduction, fusion breeding has been the ugly duckling of fusion research. This section attempts to convert it to the beautiful swan. For my entire experience in the NRL laser fusion project, I was not 'allowed' to mention it to the two leaders I worked for and with, Steven Bodner and Steven Obenschain. I hasten to add, that these are two scientific leaders for whom I have utmost respect and admiration. As a testament to their leadership, NIKE and ELECTRA were both completed on time and on budget; and a 200 Joule ArF laser was completed with no budget at all.

To begin, let's look at 'rosy' picture painted at the end of Section V. Let us start by stipulating a 2 MJ ArF laser with an efficiency of 10% as claimed in (Obenschain). But then how valid is the gain calculation? To gain some insight, let's examine the LLNL experience. While the NIF laser was being constructed, a large group there, under John Lindl (Lindl) published a theoretical calculation of the response of the target to the laser (actually, the X-rays). They considered every physical effect they could think of and did many calculations ending up with complex maps of gains in a particular parameter space [sort of like the JT-60 result in Fig. (15)]. The region of this parameter space with a gain of 10 or better was significant, and they had every confidence that they would achieve this. With the delay in the construction of NIF, the theoretical work continued. This time it was reexamined by another large group, then under the leadership of Steven Haan (Haan). They got the same result, a prediction they thought was reliable, namely that over a large region of parameter space, they would have a gain of 10.

With hindsight, we now know that these original gain calculations were optimistic by at least 3 orders of magnitude. In short there must have been many physical effects which they were initially not able accurately evaluate. Perhaps the laser plasma instabilities were more pernicious than they originally thought, perhaps there were many more energetic electrons produced, and depositing their energy in ways difficult to predict. The author studied this effect at NRL (Manheimer 2023 C) and with the URLLE group (Gonchorav). It is not simple, the earlier attempts by us and others left out some important physics.

When diagnosing their initial results, the LLNL group knew that there was more mixing of the fuel and ablator than they had planned for (Edwards). This mixing problem has persisted (Divol). (Later we give a brief digression on this.) For instance, their working numerical hydro simulations were one dimensional, that is, they assumed spherical symmetry. Using these, they could examine a large region of the relevant parameter space and calculate the gains in the entire region. To investigate mixing, they also did many, but also many fewer 2-dimensional simulations, that is hydro simulations assuming that the configuration was spherical with only variation in the r and θ dimension. The fraction of the relevant parameter space they could examine with these 2D simulations was obviously smaller and sparser. Smaller and sparser still were the few calculations they could do in 3D, i.e. variation in r, θ, and ϕ. Considering that there was almost always more mixing experimentally, than predicted theoretically, it is reasonable to conclude that they missed some important physical effects.

Then there were physical effects they knew existed but did not know how to model. For instance, the fluid (i.e. mostly electron) energy flux seemed to be less than classical, but this effect, which they knew existed, but did not understand well, they modeled with a single parameter called a flux limit.

The purpose here is not to cast blame on the LLNL group, but rather to say that the gain calculations are complex and difficult. Twenty years later, their experimental gain is still nearly an order of magnitude below their early predictions. I am nearly certain that they brought more resources to their calculations than did the NRL group.

To this author, it seems *very* likely that when the rubber hits the road, the NRL calculations, like those of LLNL, will prove to be optimistic. For instance, let us say that the maximum gain turns out to be 50 instead of 250. Then their 2 MJ laser will produce 100 MJ of fusion energy, 30 MJ of electrical energy, 20MJ of which go right back to powering the laser. No power system with that sort of circulating power ratio could be economical. And what if the laser efficiency turns out to be 7% instead of 10%? Then *all* the fusion energy produced simply goes back so the laser can power itself.

To this author's mind, it is simply prudent to have a plan in the very likely event that the case gain is 'only' 50 and the laser efficiency is 'only' 7%. In fact, this might not be as big a leap from the LLNL result as it appears. They got a gain of 1.5 in their best shot so far. But only ~ 10% of the laser energy is hitting the target, so their Q measured by energy on target is in ~15. Direct drive gives the option of getting all the laser light on the target. If deep ultraviolet ultimately works as well as X-rays in imploding the target, the gain only must be increased by about a factor of 3 from what NIF has already achieved. This then would allow for fusion breeding.

Fusion breeding is hardly a new idea. It is likely that the idea was originated by Andrei Sakharov in 1951 (Sakharov), although it may have in fact been earlier. Also, Hans Bethe argued for it in 1979 (Bethe). These are two giants of 20$^{th}$ century physics; their analysis should have received much greater attention. Hybrid fusion was studied in the United States and other places in the late 1970's and early 1980's, but was then abandoned in favor of 'pure fusion', namely using only the kinetic energy of the fusion neutrons to, let's say boil water. Much of this information on hybrid fusion is archived in a web site (ralphmoir). This site contains many early LLNL and PPPL reports, which would be difficult to access any other way. Generally, these reports considered a fusion device surrounded by a sub critical uranium or thorium blanket which provided a 'fission kick' to the power produced. These reports claimed that a subcritical fission reactor had certain advantages, particularly as regards reactor safety.

However, this 'fusion kick' does not only enormously complicate the reactor, but it is also not even necessary or advisable. For one thing, we have known how to build critical thermal nuclear reactors safely for 70 years now, so once we have the fuel, why not burn it in the way we have always burned it? Also, since the nuclear reaction produces ~ 215 MeV and the fusion reactor produces < 20, it seems likely that the fission reactor will quickly take over anyway.

In these earlier studies, fusion breeding, i.e. the use of fusion neutrons to breed nuclear fuel for other free-standing fission reactors was hardly ignored (Moir 1982 A&B, 2013), (these references called this fission suppressed hybrid fusion). However, this was certainly not emphasized either. This author sees the main justification for fusion breeding as combating a potential shortage of nuclear fuel. Furthermore, using fusion simply to breed fuel for current, and likely future stand-alone fission reactors fits in much better with current, and likely future nuclear infrastructure. In short, it is not only possible, but likely, that fusion breeding could fill a pressing future need. As discussed in Section II, fusion breeding is the only breeding approach than can refuel 'stranded' thermal reactors that otherwise would be 'out of gas'.

In fact, if fissile nuclear fuel were sustainably available, a competition between a fusion reactor and a fission reactor, say an LWR is almost certainly unwinnable for the fusion reactor. However sustainable fuel from either mines or the sea is hardly a sure thing, in fact it is almost definitely *not* a sure thing. The relevant competition is between fusion breeding, fast neutron breeding, and thermal neutron thorium breeding. This is a competition for which fusion breeding as many advantages if we can pull it off. Most likely the optimum will be some combination of the 3.

The first issue with fusion breeding is that the fusion reaction produces only a single 14 MeV neutron. It is necessary to produce tritium from Lithium, so one might think there would be no neutrons left to breed $^{233}$U from thorium. However, as mentioned in Sec IIIA, this is not the case; the saving grace is the high energy of the fusion neutron. By a process called spallation, the neutron can collide with other nuclei and produce additional neutrons. Lead is a good target for neutrons with energy above 7 MeV. This reaction is:

n + Pb → Pb + 2n (-7 MeV)

An even better neutron multiplier is beryllium, which needs only a 2.7 MeV neutron to produce an additional neutron. This reaction is:

n + Be → 2He + 2n (-2.7 MeV)

It seems that a single 14 MeV neutron, in a target of $^{9}$Be, could, in theory at least, produce as many as 4 or 5 neutrons as it traverses a mostly beryllium target, ultimately producing that many $^{233}$U's and tritons. The cross sections for some of the most relevant breeding reactions are shown in Fig (22).

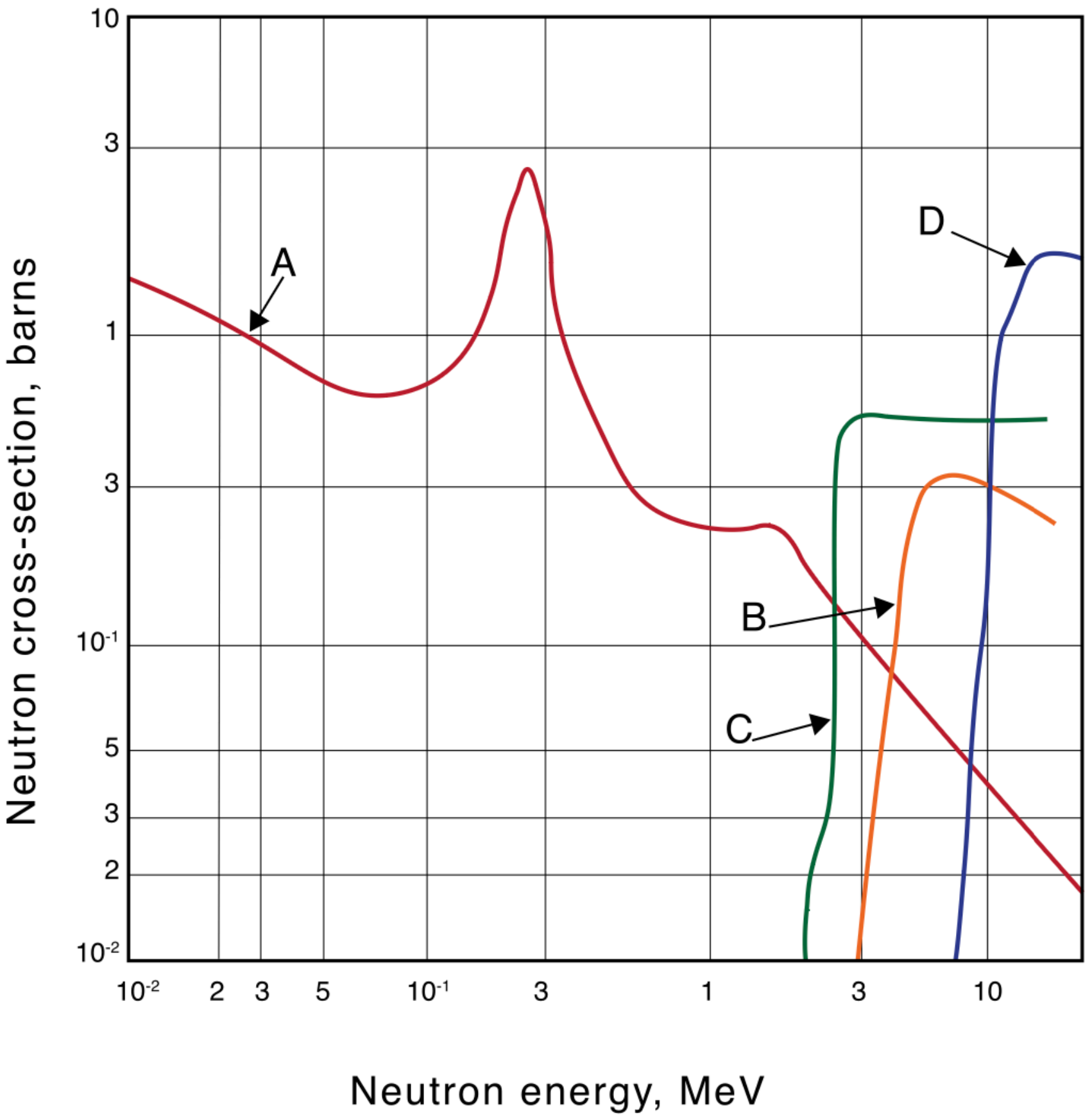

Figure 22: Collision cross sections for several of the most important reactions for fusion breeding. The red curve (A) is the cross section for tritium production from $^{6}$Li; the orange curve (B) for tritium production from $^{7}$Li, a process that preserves the neutron; the green curve (C) the cross section for a neutron spallation collision with $^{9}$Be, and the blue curve (D), neutron spallation from Pb.

All DoE labs study have studied these processes with the use of what are called Monte Carlo codes. They start with many 14 MeV neutrons and study their paths in the material specified. As they collide with nuclei in the material they are in, they both lose energy and generate many daughter nuclei, which the code also follows. Once all the particles have zero energy or an energy below some minimum, the simulation stops, and enumerates how many daughter particles of are produced and the total energy liberated. Some examples of the daughter particles produced by a single 14 MeV neutron as it slows down in various materials are shown in Table 1 (Moir 2013)

| Medium | Product atoms | Energy released (Mev) |
|---|---|---|
| $^{232}$Th + 16% $^{6}$Li | 1.3 $^{233}$U + 1.1 T | 49 |
| $^{9}$Be + 5% $^{6}$Li | 2.7 T | 22 |
| $^{9}$Be + 5% $^{232}$Th | 2.66 $^{233}$U | 30 |
| $^{7}$Li + 0.8% $^{232}$Th + .02% $^{6}$Li | 0.8 $^{233}$U + 1.1T | 17 |

TABLE (1) Product atoms and energy released by a 14 MeV neutron impinging on various homogeneous materials. Notice that all of these reactions are exothermic, so they increase the power of the fusion reactor by a factor (called M) above the purre nuetron power. Here we treat M as roughly 2, so to produce a 3GWth reactor, one needs only to produce 1.5GW of neutron power, rather like the desgn power of the original Large ITER.

The fusion blanket, then is not only a heat exchanger, but it has some beryllium, lead, or $^{238}$U for neutron multiplication, and then has some Li and some $^{232}$Th which breed tritium and $^{233}$U. The design of these blankets is complicated, and little work has been done on fusion breeding. Accordingly, this paper hardly sees the published designs so far as optimum.

The nuclear process is that a thorium nucleus absorbs a neutron, becoming $^{233}$Th. However, this nucleus is unstable, and nearly immediately expels an electron, moving one unit up the periodic table to become $^{233}$Pa (protactinium). This is also unstable, this with a half-life of about a month. The nucleus expels another electron and becomes $^{233}$U, which is stable and is a perfectly good fuel for a thermal nuclear reactor. A diagram of the complete reaction from fusion plasma to $^{233}$U is illustrated in Figure (23).

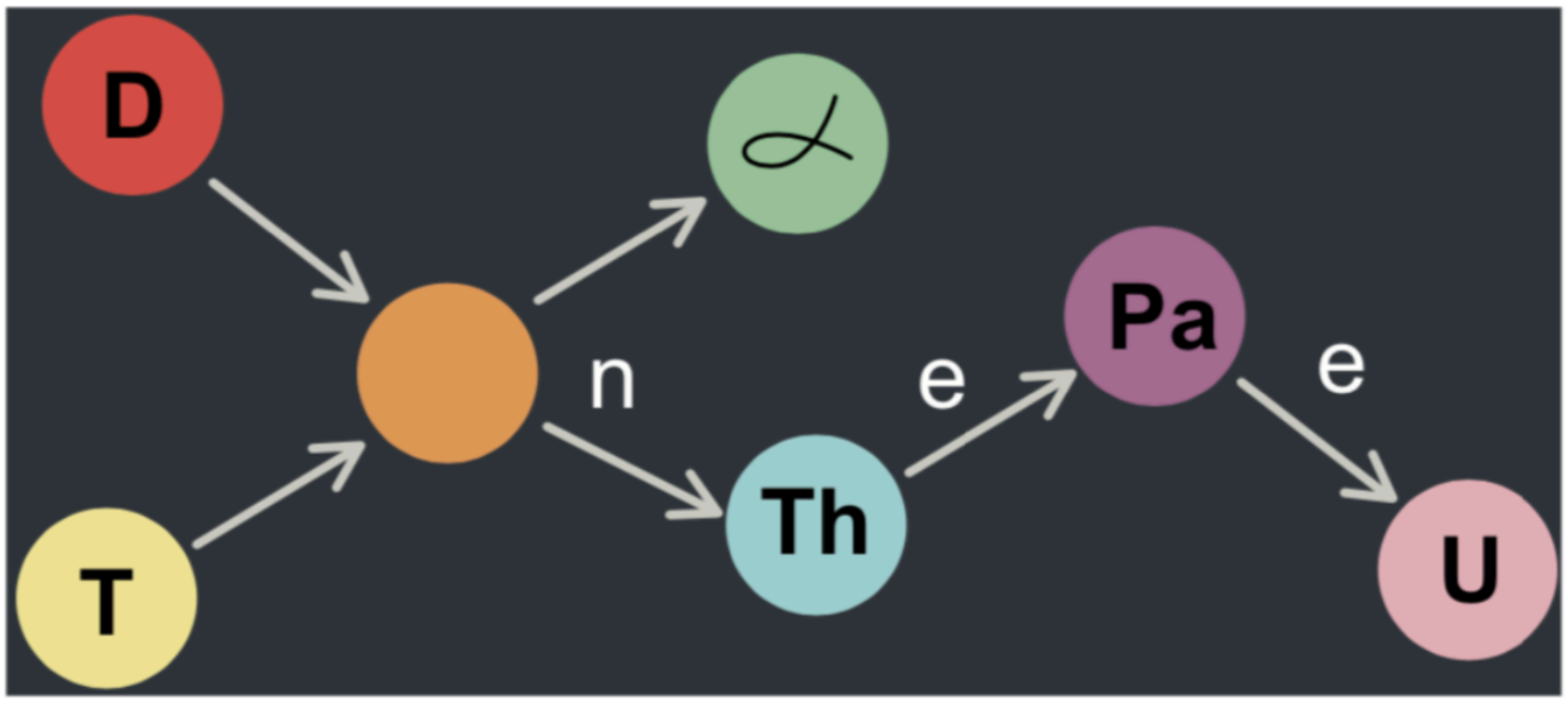

Figure 23: A schematic of the decay process where a fusion neutron is absorbed by a thorium atom, setting into motion a decay process ending up with $^{233}U$, a perfectly good reactor for a thermal nuclear reactor.

In fusion one neutron is needed to breed the tritium from lithium, so in either case, after the spallation neutrons are produced, one or two neutrons are available for other purposes. Of course, in either case there are losses, so probably somewhere between half and one neutron per reaction is available for breeding $^{233}U$ from $^{232}Th$, or $^{239}Pu$ from $^{238}U$. However, the fission reaction produces ~ 215 MeV, while the DT fusion reaction produces most of its energy as a 14 MeV neutron. Hence for reactors of equal power, a fusion reactor generates about 10 times more neutrons, and therefore breeds about 10 times more nuclear fuel than a fission reactor does. Hence it takes two fission breeders to fuel a single thermal reactor of equal power at maximum breeding rate. However, due to the additional neutrons per reaction, and the lower energy of the fusion reaction, a fusion breeder can fuel 5-10 thermal reactors of equal power. In other words, a fusion reactor is neutron rich and energy poor, while a fission reaction is energy rich and neutron poor, a perfect match.

Notice that the relative energies of the two nuclear reactions, the breeder and the bred, are the keys to the breeding capabilities of the various reactions. The neutron energy from the fusion reaction is 14 MeV, and the neutron absorbed by the fertile nucleus produces fission energy is ~ 215 MeV, so the fusion neutrons can breed fuel for ~ 5-10 thermal nuclear reactors of equal power. A fission breeder, and the thermal reactor it is fueling, each have a basic energy of ~ 200MeV, so a fission breeder can fuel 0.5-1 thermal reactor of equal power. On the opposite end of the spectrum, is a CANDU reactor producing tritium. The basic CANDU nuclear reaction, again is ~ 200MeV, but the reactor it is attempting to breed for is a fusion reaction which gives only 20 MeV. This is why even the world's 40 CANDU's cannot fuel even a single fusion reactor. Getting tritium for the first commercial fusion reactor is as significant obstacle, which will be discussed in the next section.

In a pure fusion reactor, the liner can be either a solid or a flowing liquid. If it is a solid, the tritium can be recovered only when the liner is removed, probably every year. This alone wastes 4% of the tritium produced. The complex geometry of a tokamak or stellarator makes a solid liner very difficult to remove. If the liner is a flowing liquid, flowing in and out of the fusion region, the tritium can be removed as it forms. Generally, the earlier reactor designs had the fluid flowing through a variety of pipes. As we will see, laser fusion opens the possibility of a fluid liner which has a free surface, where pipes may not be necessary.

For a fusion breeder, the only option for the liner is a flowing liquid. If it were a solid liner, $^{233}$U and $^{233}$Pa would accumulate in the liner, and in the incoming fusion neutron flux, would certainly burn more and more as more of these nuclei are stuck in the liner, the power would rapidly continually increase until the liner is destroyed.

The fluid blanket is generally assumed to be a molten salt, and FLiBe ($Li_2BeF_4$) is the most common example. It has 2 nuclei of lithium for every one of beryllium for breeding tritium. The Li breeds the tritium, and the Be is a neutron multiplier. Also, uranium, protactinium and thorium are all soluble in it. The thorium can be introduced at the input, and protactinium can be extracted at the output. Also, FLiBe has a melting point of 462$^o$C and a boiling point of 1430 $^o$C, so it can be an excellent heat exchanger. For instance, if the FLiBe enters the fusion region at a temperature just above the melting point, and exits just below the boiling point, that would give rise to a potential Carnot efficiency of nearly 60%. Reference (Moir A) gives an example of a preliminary design of a FLiBe blanket, with Thorium dissolved in it and calculated the reaction products with DoE Monte Carlo codes. They calculate that each 14 MeV fusion neutron will produce 1.1 Triton, 0.6 uranium 233 nuclei, and the total energy released will be ~25MeV.

We think in terms of a fusion reactor producing ~ 1.5 GWth neutron power like the original large ITER. The breeding reactions increase this to ~ 3GWth. Add the alpha power and the reactor produces a total of ~ 3.3 GWth, just like a conventional reactor. As the14 MeV fusion neutrons produce 1.5 GW, the 0.7 $^{233}$U's from each neutron, produces nuclear fuel at a rate of ~15 GW or more, about enough to fuel at least 5 thermal nuclear reactors of equal power.

Let us see how this works out for our example of a 7% laser with a target gain of 50. The 2 MJ laser produces ~100MJ of neutron energy, which produces ~1000 MJ of fuel. If the laser is pulsed 15 times per second, this would breed ~ 15 Gigawatts of fuel power, enough for about 5 LWR's of equal power. It is the difference in energy between the individual reaction in the fusion reactor, and that in the fission reactor which is mostly responsible for this large gain.

## VII. Some digressions

This section digresses on several aspects of fusion which seem to be important, but which appear to have received little or no attention. The goal here, obviously is not to present complete solutions, but to give preliminary thoughts and suggest future work.

### VII A: A Different kind of reaction chamber allowing for a free surface liquid flow:

Many have argued that even for pure fusion, the first wall of a fusion device should be a flowing liquid. For fusion breeding, this appears to be the only choice. As there is no lattice structure in the fluid, one does not need to consider things like how many atoms are moved from their lattice site, after so and so many interactions with energetic fusion products. Thus, the liquid 'self-anneals'. Furthermore, one does not have to wait until the liner is replaced to recover the bred tritium and/or $^{233}U$. The liquid flows first into and then out of the fusion reaction chamber to a chemical plant where the various bred quantities are separated out. The most obvious advantage would be if the liquid had a free surface facing the fusion plasma. If this is not possible, the liquid would have to flow in pipes, meaning that for instance that the 14 MeV neutrons would have to first interact with the pipe material before penetrating the fluid. In time this would damage the pipes, and perhaps degrade in some way the breeding processes in the fluid.

Look at the schematics of a tokamak, Fig. (9) or a stellarator, Fig. (10). Getting a solid liner in or out to recover the tritium or $^{233}U$ looks like a very tough job. Getting a free surface liquid to flow smoothly along the very irregular wall looks to this author nearly impossible. Even getting fluid flow in pipes to do this looks especially tough. For instance, Figure (24) (ralphmoir) shows an earlier design of the pipe system for a fusion breeder based on a magnetic mirror fusion configuration, a which is much simpler than either a tokamak or stellarator.

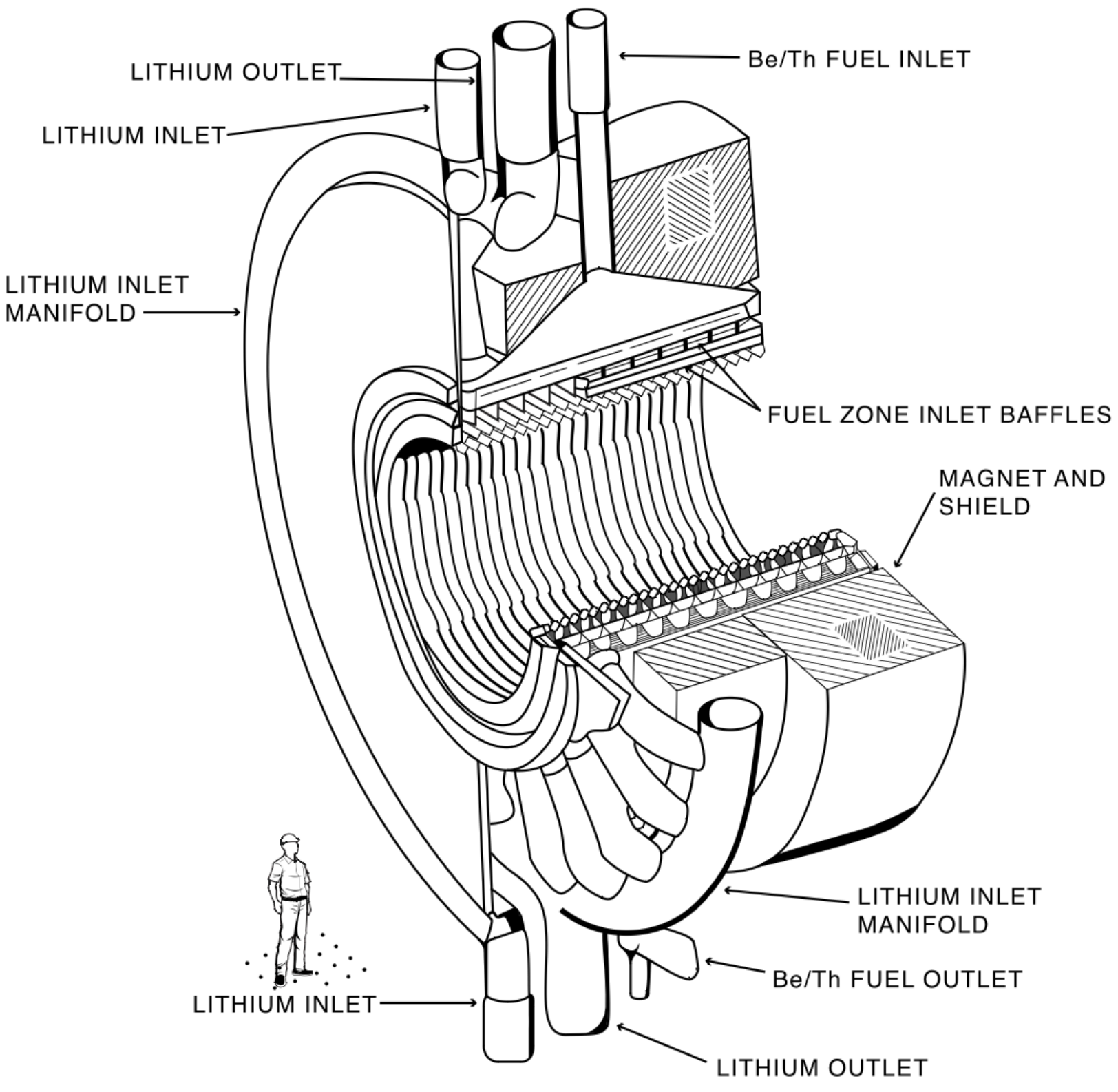

Figure 24: A schematic of an earlier concept of the pipes for the liner fluid flow for a fusion breeder, but for a magnetic mirror reactor, a much simpler design than what would be required for a tokamak of a stellarator.

Notice that for a laser fusion configuration in a spherical target chamber, Figs (11 and 19), it would be very difficult to get the pipes in, and at least to this author, a flowing free surface boundary the spherical configuration, with many holes in it seems just about impossible.

However it is worth noting that laser fusion, which produces a point source of fusion products, does open up the possibility of of a liquid blanket with a free surface. Instead of a sphereical reaction chamber, one could use a segmented cylinder with liquid flowing down the sides, as shown in Figure (25).

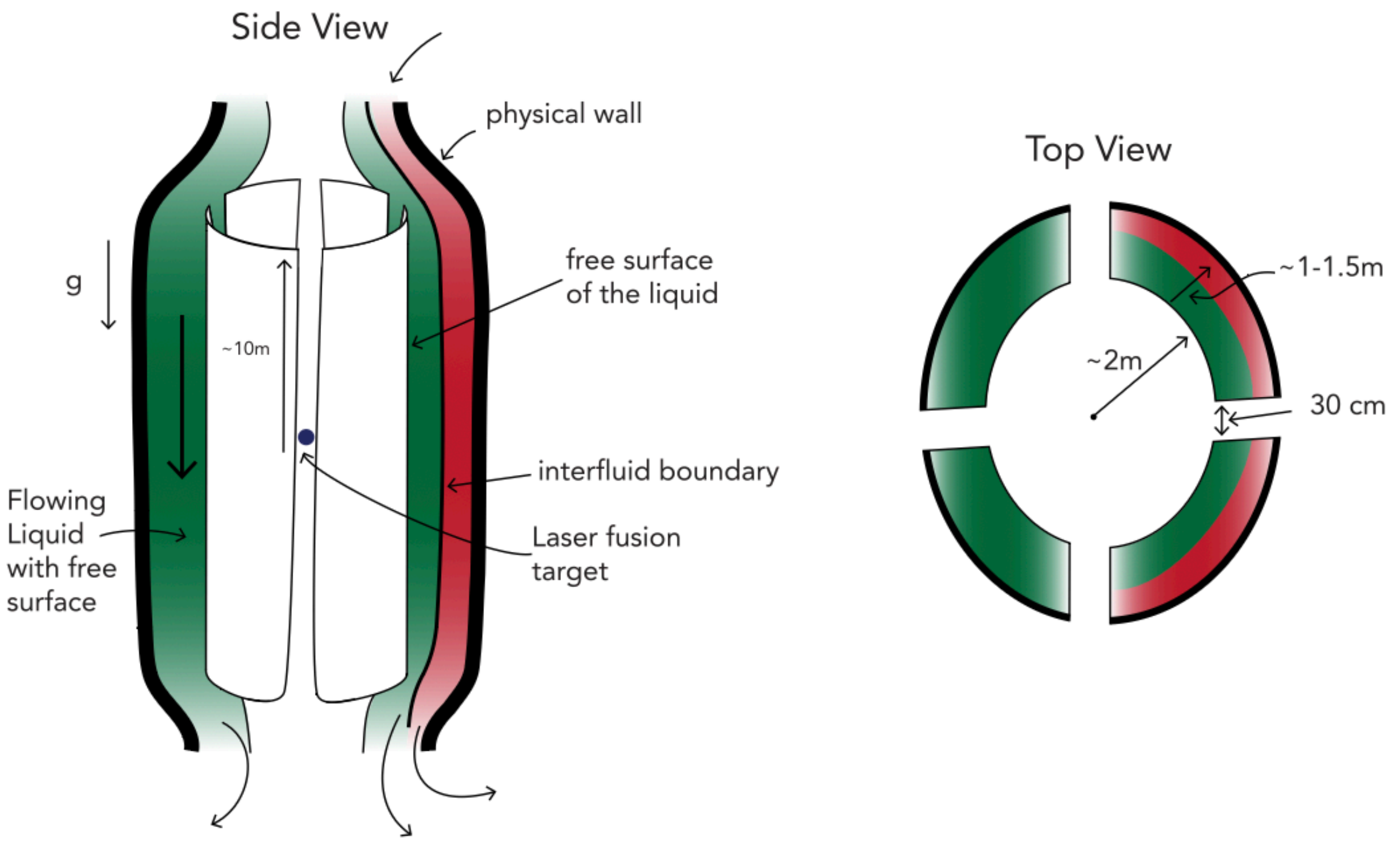

Figure 25: A schematic of a liquid blanket with a free surface flowing down a segmented cylinder. The measurements shown are a 2-meter radius to the wall, and a total length of 20 meters, i.e. a half-length of 10 meters. The side view is on the left. The flow going down the left-hand side of the cylinder, the green fluid, is assumed to be a single fluid, on the right 2 fluids, green and red, with no physical boundary between them. With the measurements shown here, the flowing liquid will absorb ~85% of anything emitted by the laser fusion target. With the 4 slots as shown and the open top and bottom, the laser could either be focused on the six faces of a cube, or on the 8 faces of an octahedron.

If for some reason the free surface flow is not viable (for instance, turbulence, evaporation…), it is simple enough to use pipes in this configuration. The pipe structure would be much simpler than the pipe structure in the spherical target chamber with many hole in it like that shown in Figures (11 and 19); and it would be much, much simpler than the pipe structure in a tokamak or stellarator. If fact if pipes were used, much of the vertical slots could be closed off.

Figure (25) gives an example of the vertical flow of a free surface liquid down vertical wall in 4 segments. Another possible configuration is to have vertical the cylinder broken up into 5 segments. In this case the laser light accesses the target from the top and bottom, as well as 2 beams coming in through each of the 5 slots. The 12 beams would be focused on the 12 faces of a regular dodecahedron. Not only is the dodecahedron a better approximation to a sphere than is a cube or an octahedron, but its pentagon face is a better approximation to a circle, the natural focal cross section of the laser beam, than is a square or triangle.

VII B An approach to reduce the mixing

Mixing of the ablator and fuel has been a constant headache in the LLNL implosion experiments. For instance, in their 2013 Phys. Plasmas paper (Edwards) on what was to be the summary at the final time of the NIC campaign, they mentioned this as one of their biggest issues. Their target mass was about 200 μg of DT. LLNL had estimated that with about 0.1 μg of ablator mixing with the target, they would be okay. However, their measurements indicated that about 3–4 μg of ablator were mixing with the target, 30–40 times the acceptable level! This problem has persisted for nearly the next decade. Just before they got a second working implosion, which they presented at the 2022 APS-DPP meeting in Spokane, they did not think they were going to be successful. Laurent Divol's abstract (Divol) for his plenary talk, submitted a few months before the meeting, emphasized this mixing as a source for their current failure. Quoting from his abstract:

"Low mode hot-spot asymmetry predominantly explained by residual laser power imbalance and capsule asphericity. The resulting shell asymmetry lowers the stagnation pressure and confinement. – Mix induced by the fill tube and ablator imperfections. Visible ablator jets and meteors in the DT fuel increase the radiative cooling".

Well, whatever their problems during the preceding year, they managed to get around them, enough so that by the time the meeting rolled around they had a spectacular result to report.

Nevertheless, the fact remains that asymmetries and mixing have proven to be a big problem over at least a period of a decade. While naturally worrying about being 'a fool rushing in where angels fear to tread' (I have never participated in designing a target); and with all humility, this article suggests something which may help the problem. Namely use a target that is DT and only DT. Figure (26) shows a picture of such a spherical target. Like the target in Fig. (12), it is gaseous DT in the middle, surrounded by an ablator of DT ice. Around the DT ice target is a very thin shell, perhaps plastic. Its purpose is only to hold the target together as it is being stored and moved. It would be blown away by a brief laser prepulse, and when the actual laser pulse strikes the target, it would be illuminating a target of only DT. If nothing else, the 'visible ablator jets and meteors in the DT fuel' could no longer 'increase the radiative cooling' since no higher Z impurities would be introduced.

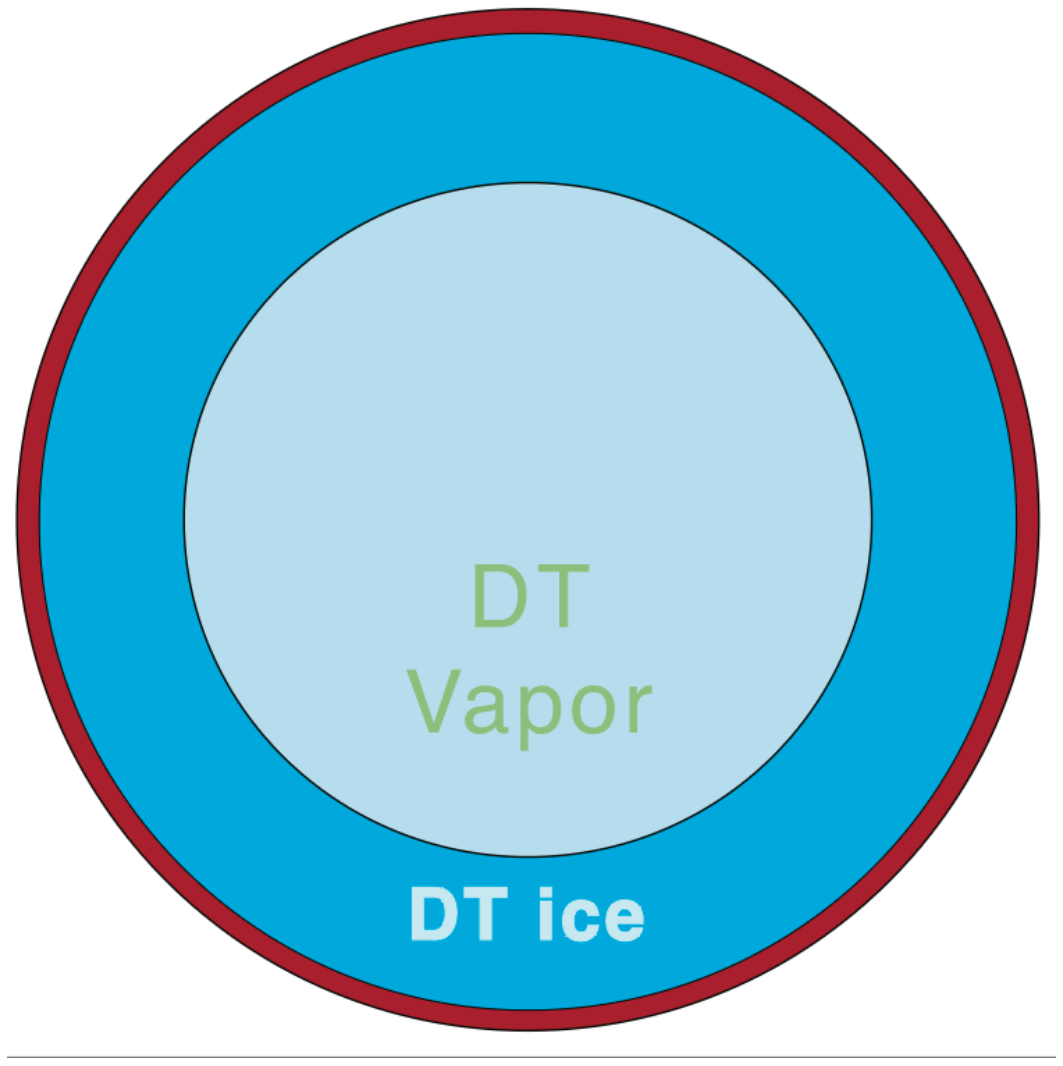

Figure 26: A schematic of a potential laser fusion target, proposed to minimize effects mixing the ablator and fuel. The inner pale blue part is a DT gas; the darker blue is DT ice, and the red part is a shell only to hold the target together during storage and transport. As soon as the laser prepulse strikes, this blows away leaving a target of pure DT. Any ablator which mixes with the fuel, simply becomes the fuel and introduces no higher Z impurities in the fuel.

VII C: How fusion might begin

In Section IV we showed that there is most likely sufficient tritium produced by existing and future CANDU reactors for the fusion research project to proceed. But then what? Once a fusion economy is established, one can acquire the tritium by breeding in the fusion reactor as discussed. But how does one get the first fusion reactor going, and what about the second if it takes all the thermal nuclear reactors just to supply the first? As we have seen, it would take the 40 CANDU's a year to produce enough tritium for about 2 week's or a month's fuel for a fusion reactor of equal power.

Here we discuss this, a serious issue which will take proper organization to pull off. This paper sees the tritium supply, from now until a fusion economy, proceeding in 3 stages. First, there the experimental stage, the stage we are in now given the LLNL successful fusion experiments. In a little over 10 years, if ITER proceeds without further delay to DT plasmas, the experimental stage will demand a much larger supply of tritium, which will only be exacerbated if more ITER class magnetic fusion research devices are brought online. This demand can most likely be met by the world's CANDU reactors without making any changes to what the reactors are already doing. Of course, it would depend on how many ITER class magnetic fusion research reactors are running, and how long and intense the research phase must be before a pilot plant could be built. Second, there will be a stage of building the first one or two fusion pilot plant reactors which run most of the time. These will make much greater demands on the tritium supply, which the CANDU's cannot satisfy. As we have seen, a 3 GW reactor will burn ~$10^{21}$ tritons per second. The 40 CANDU's together produce ~$4x10^{19}$ per second. This new demand can be met only by the nation's and/or world's supply of light water reactors. However, these reactors will have to use a different type of fuel rod. Third, and final, there is the demand for tritium for the world's supply of fusion reactors. Fortunately, this is a demand that can be met by fusion reactors themselves, but they will have to use a blanket and a reaction quite different from the ones discussed so far. This blanket will allow for an exponential growth of fusion reactors in time, ultimately filling the need.

Stage 1: The research stage: As we have seen in Section IV, the CANDU reactors are producing enough tritium to support the DT fusion research. This is certainly true for laser fusion research with DT. This needs very little tritium on each shot. It is almost certainly true for ITER scale research with DT, which needs considerably more tritium, but does not seem to be beyond what CANDU's can generate. If there are more ITER scale research magnetic fusion reactors, and a great deal of research with these reactors is necessary, there may be a problem.

Stage 2: Setting up the first one or two commercial pilot plant reactors: We have seen in Section IV that all the CANDU reactors in the world cannot even come close to fueling a single 3 GWth fusion reactor. Some other means must be found. Here is where the world's LWR's come in. The fuel rods of a light water reactor can be made with some small fraction of $^6Li$ inside. In fact, the United States is already doing this at the TVA Watts Bar reactor in eastern Tennessee (Caffey, Tritium). The purpose here, is to produce tritium for the American nuclear deterrent. There are about 100 American LWR's. Each one produces ~ 1 GWe, for a total of ~ 100 GWe. Let's say that each one is fueled with enough lithium 6 to produce 1 T nucleus for every 10 nuclear reactions. Most likely this would mean about a 10% reduction in the power produced. Since the fission reaction produces about 10 times as much energy as the fusion reaction, and there are 10 times as many fission reactions, all 100 LWR's together, in this configuration, would produce enough tritium for a single 1 GWe fusion reactor. This way, fusion reactions could get a start, but clearly there is no way that a fission economy could support a full fusion economy.

Stage 3: After we have 1 or 2 fusion reactors, fueled by the American or world's LWR's: The key is having the fusion reactor produce maximum tritium, not economic power. As we have seen in the previous section, with a properly designed blanket, one with beryllium and diluted with 5% $^6Li$, it is possible that each 14 MeV fusion neutron will produce 2.7 tritium nuclei and 22 MeV energy to use ultimately produce electricity. Hence if the reactor designers can pull this off with a flowing liquid blanket, which they can rapidly treat in the chemical separation plant, they will produce enough tritium to fuel not only their own reactor, but 1.7 other reactors as well. Each new reactor brought online, of course can do the same thing, allowing, the supply of tritium (and number of reactors) to increase exponentially with an e-folding time of about a year. Once there is enough tritium to start all reactors, the problem is solved. As additional fusion reactors wish to go online, the existing fusion reactors can always switch to the beryllium lithium blankets to get tritium for new reactors. To the author, this reaction and blanket seem to be a 'gift from God'. Without this, or a reaction and blanket like it, it is difficult to see how fusion, could ever become more than a curiosity, and how commercial fusion could ever be more than a pipe dream. With it, a fusion economy for world development appears possible.

The only question is what fluid does one use in the blanket? Clearly FLiBe is not a satisfactory fluid, since it has twice as many lithium nuclei as beryllium, where this process requires that the blanket have only 5% lithium. One possibility is to use $LiCl_2$ as the molten salt. It melts at a temperature of 405° C and boils at 520°, so it is obviously not a good fluid for a Carnot heat cycle. But this is not its purpose; its purpose is to produce tritium for itself and other reactors. Of course, one needs lithium as well, and there are several possibilities. One could add enough FLiBe so that atomic ratios of Li and Be are proper. If the FLiBe is to be a fluid, it would mean operating at higher than the minimum temperature for $BeCl_2$. Also, one could simply add lithium chloride or fluoride (or FLiBe) in its solid phase and mash it up into tiny particles to be carried along with the flow, sort of like hazy ale. Even in a worst-case scenario where no fluid blanket can do the job, it is possible to use a solid alloy of beryllium and lithium 6, as was assumed in (Moir 1982 B, and 2013) While this would obviously slow things up as compared to a fluid blanket, in a reaction chamber like that shown in Fig (25), the simple geometry would allow solid blankets to be removed and inserted quickly as compared to say a tokamak or stellarator. There seem to any number of possibilities.

This reaction, which allows for an exponential growth in tritium supply in time, with an e-folding time of about a year, appears to be one of the neglected, but vital keys for fusion development.

VIID: What if the NRL calculations of Q=250 and η =10% prove to be correct?

At NRL and other places, the argument against fusion breeding claimed emphatically that there is no need for it, pure fusion is possible and achievable. After all the argument made here for fusion breeding, is that the NRL calculations of laser efficiency and target Q were most likely optimistic, just like they were at the analogous program at Livermore. But what if they turn out to be correct? Does this put fusion breeding out of business? This author's answer is an emphatic **NO!**

In (Manheimer 2023A) this author presented an analogous argument for an ITER based fusion or fusion breeding device. These arguments were based fundamentally on published and assumed cost of an ITER type device, and on the price of electricity. It made the case that a Q=10, 3GWth, $25B ITER type device could evolve to an economic breeder but could not possibly be a stand-alone economic power supply. But what about a Q=40, 3GWth, $4B device? This could certainly be a good, economic stand-alone power source. However, with very little extra effort, it could fuel many thermal nuclear reactors, with fuel cost nearly 'too cheap to meter'. Why would anyone not do this?

This author has not seen any cost estimates for a laser fusion-based power supply, so we will use a somewhat different logic, one which make a variety of reasonable assumptions about the various costs, deduced from an assumed end point cost for a kWhr of electrical energy. Let's say that the NRL calculations of Q (i.e. 250) and η (i.e. 10%) turn out to be correct. Let's say that with this 3 GWth (1 GWe) power supply, the utility can sell power to its customers at 12 cents per kWhr. Let's also assume also that this cost can be further broken up into half the cost for the power supply, and half for other costs, stringing up the wires, repairing storm damage…… Hence the power cost from the laser fusion power station is 6 cents per kWhr.

We also make the reasonable assumption that that the cost of the power supply scales as the average power of the laser. Well, then consider putting a fusion breeding blanket around the target, only to double the power of the fusion system at virtually no extra cost. Then the laser fusion system would only need half the average power, and half the cost to generate 1.5 GWth of fusion power instead of 3 GWth. Hence for incorporating this blanket alone, the cost of power at the end point is reduced from 12 cents per kWhr to 9 cents per kWhr.

But this breeding blanket does not 'only double the power of the fusion system at virtually no extra cost'. It also produces fuel for many thermal nuclear reactors at nearly no extra cost. Again, nuclear fuel becomes nearly 'too cheap to meter'. To this author, it seems inconceivable that our descendants will not choose this option, especially where there may by then be a crying need for nuclear fuel for a worldwide multi trillion $$$ investment in thermal reactors which might be 'stranded and out of gas'. However, whatever they choose, it is not our decision to make. It will be made by our grandchildren and great grandchildren. Getting the best, most economical fusion system, as quickly as possible, is the biggest favor we can do for them. There is absolutely no downside in making more pessimistic assumptions, and planning for a fusion breeder, instead demanding only pure fusion.

## VIII: Fusion Bureaucracy 101 (a course I'd probably fail)

When making recommendations publicly for the fusion program, as this article does, one must realize fusion is in a strait jacketed bureaucratic environment. Like it or not, this is the environment in which the fusion effort lives. In the US, there is the fusion budget for nuclear security and stockpile stewardship. This supports LLNL and NIF, as well as programs at other DoE labs like Sandia. As this paper is concerned with energy for the civilian sector, a very different goal from nuclear security, it seems to this author that it remains entirely separate from that for nuclear stockpile stewardship.

The yearly budget for magnetic fusion energy for the civilian sector ~$740M. Of this, ~$220M is our yearly commitment to ITER. ITER is an important international project, now ~ 75% built. It is important that this project continues. It has a huge fusion infrastructure in place, and will almost certainly make important contributions to fusion, whether the final fusion device is a tokamak or something else. While this author favors laser fusion, nobody's crystal ball is perfectly clear.

Then there is the remaining $520M per year to support the domestic magnetic energy fusion program, efforts like the General Atomics tokamak research using their tokamak D3-D, and the Princeton Plasma Physics Lab, as well as a recent investment into potential private sector fusion development (DoE). It is here that this paper argues for major bureaucratic changes, changes on the order of several hundred million $$ per year. Obviously, many feathers will be ruffled, and those on the losing end will fight hard to keep their money and power. Nevertheless, within the DoE fusion energy department, yearly $$ on order of several hundred million, let's say $300M should be reprogrammed from magnetic fusion to inertial, i.e. laser fusion for civilian energy. This portion of the American Department of Energy program could be split up into two separate

branches, say ~$300M for laser fusion, and ~$200M for magnetic fusion. Needless to say, several American magnetic fusion efforts, good programs, would have to be cancelled. But no government program is guaranteed eternal life. As needs change, so must programs be put in place to meet these changes. Furthermore, at this juncture, the rest of the world has many continuing domestic magnetic fusion programs, tokamaks and stellarators in Europe, Japan, China, India, Russia…., to say nothing of an American private 'fusion start up' or two. If the US mostly pulls out of tokamak research, there are many who will continue it. At this point at least, the US is the only country with a credible program on laser fusion.

The alternative might be to find hundreds of millions of new $$ to add to the American government's fusion budget to support laser fusion for civilian energy. However, in view of the American government's gigantic budget deficit, this seems very unlikely. Given fusion's 60-year effort, with a payoff still decades away, this does not seem a good time to ask the beleaguered American taxpayer for more.

Furthermore, it seems to this author that the American magnetic fusion efforts have hit something of a brick wall. They all seem to be waiting for results from ITER, probably ~ 20 years from now, if there are no further delays (Kramer A). Hence, there is a strong argument for replacing much of the government sponsored magnetic fusion effort in the United States with an excimer laser direct inertial fusion energy (IFE) effort, while still honoring its commitments to ITER.

For instance, in searching the PPPL web site, there are two main experimental programs the lab is now involved with. One is helping the Japanese with their tokamak JT60-SA. The other is rebuilding their spherical tokamak (ST) which has been down for quite some time because of a broken coil. They claim this latter project is now 76% complete, and it will investigate liquid metal linings. However, ST's almost certainly cannot lead directly to an economical fusion reactor, as it is unlikely that the center post can withstand the intense neutron flux in a reactor and remain superconducting. In other words, PPPL is now a lab in service to other labs or is doing projects not in the main line of fusion research. It is no longer a leader of the fusion project as it had been from 1970-2000.

Switching to direct drive laser fusion is most likely the best way PPPL, or any other American DoE lab; can reclaim American leadership in fusion. This takes a change not only in the direction of the science project, but also a change in the bureaucratic structure supporting it. Instead of the US Department of Energy branch on civilian energy, supporting only magnetic fusion, that department should be broken into 3 subgroups. The first supports ITER with the existing $220M yearly budget. The remaining parts ($520B) should be split into two subgroups, one supporting laser fusion with a budget of ~ $310M, and a another supporting magnetic fusion with a budget of ~ $210M.

Also there should be a DoE lab focused on laser fusion for civilian energy. It might be an existing DoE lab with a goal modified to support laser fusion, or it might be a completely new lab. This lab will cooperate somewhat and compete somewhat with LLNL. However, since the goals of the two labs are so different, this author believes that cooperation will be maximized, and competition minimized.

Bureaucratically, it is not easy to see how to pull this off. As stated in the Introduction, it looks like a slog through a quicksand bog a mile wide and a mile deep. However, this author firmly believes that it must be done, and the sooner the better.

The establishment of a separate DoE lab for excimer laser driven fusion is hardly the only bureaucratic challenge facing the fusion effort between now and when fusion becomes operational. As we saw in the last section, there are many future tasks for which a bureaucracy will have to be installed to manage. Let us look at some of these tasks. When ITER becomes operational, it might not be so simple to get the tritium from the world's CANDU reactors. This must be managed in some way for the duration of the experimental programs, and before fusion power plants are online. When fusion research becomes advanced enough that it is prudent to set up a pilot plant, then we must manage the transition of ~ 100 LWR's to use fuel rods with $^6$Li, impurity so that they produce enough tritium to fuel these one or two initial fusion reactors for a year or two. Once these reactors are established, they must use blankets which manage tritium production, not blankets which maximize power. They must use blankets which allow for the growth of the tritium supply to rapidly increase exponentially in time until enough is produced. Finally, when enough tritium is produced, it will be the time for the fusion reactors to fuel themselves with the necessary tritium, in pretty much the way that the fusion project has always suggested. The conventional fusion program knows how to keep fusion going, just not how to get it started. To do the complete job takes not only wise and capable scientists and engineers, but also wise and capable bureaucrats.

## IX: The energy park

Since 2004, every article this author has written on fusion breeding has ended with a section on 'The Energy Park', and this one is no different. Large parts of this section are taken from Section VIII of (Manheimer 2022B), which has been published by the author open access. The energy park is a proposed key element of an energy infrastructure to supply tens of terawatts to the world. It uses the fact that a fusion breeder can breed fuel for about 5 (LWR's) of equal power, and each year an LWR discharges about 1/5 of its fuel as plutonium and higher actinides.

As discussed earlier, a reactor like a light water reactor discharges waste in 3 categories. First there is uranium, which is separated out in the energy park for reuse. Second there are the highly radioactive fission fragments, like the krypton and barium shown in Fig (2). These have a half-life of ~ 30 years, and they have no proliferation risk. Some have economic value and would be separated out and sold. The rest would be stored in pools or encased in some way. They would be stored for 300-500 years. This is a time scale human society can reasonably plan for. After this, they would have decayed sufficiently that they could be properly diluted and released in some way into the environment.

Third are the transuranic elements, especially $^{239}$Pu, which has a half-life of 24,000 years. The energy park proposes to burn the discharged actinides with a fast neutron reactor like the IFR.

This is different from the French approach, which recycles these actinides into fuel for thermal reactors. The thermal reactor also creates additional actinides, of constantly higher atomic number. A portion of the nuclear wastes is burned, but another portion becomes a more and more complex stew of higher and higher Z actinides. The advantage of fast neutron reactor is that one time through, it burns all actinides equally as the cross sections shown in Section II have shown. There is no endless recycling, a single burn will take care of all the actinides.

This series of papers have invariably assumed that the transuranic products of the thermal reactors must be rendered harmless. The alternative is burying them somewhere and creating what amounts to a 'plutonium mine', like Yucca Mountain. This would plague society for half a million years or so. This is an immoral burden to lay upon our descendants, hence the need for the fast reactor in the energy park.

One envisions an energy infrastructure where there is one fusion breeder to supply fuel to about 5 thermal reactors like the LWR or more advanced thermal reactor, and one fast neutron reactor to burn the 'waste' actinides. If a more advanced thermal reactor is developed, one which burns more deeply into the fertile material, and produces fewer actinides in its waste, the fusion breeder and fast neutron reactor could perhaps service as many as 10 more advanced thermal reactors.

The fast neutron reactor could be something like the Integral Fast Reactor (IFR), developed by Argonne National Laboratory. It ran successfully at 60 MW for years before it was disassembled. It could run on any actinide and could run in either a breeder or burner mode.

A schematic of the energy park is shown in Figure (27). Most of the elements of the energy park are available today, only the fusion breeder needs full development.

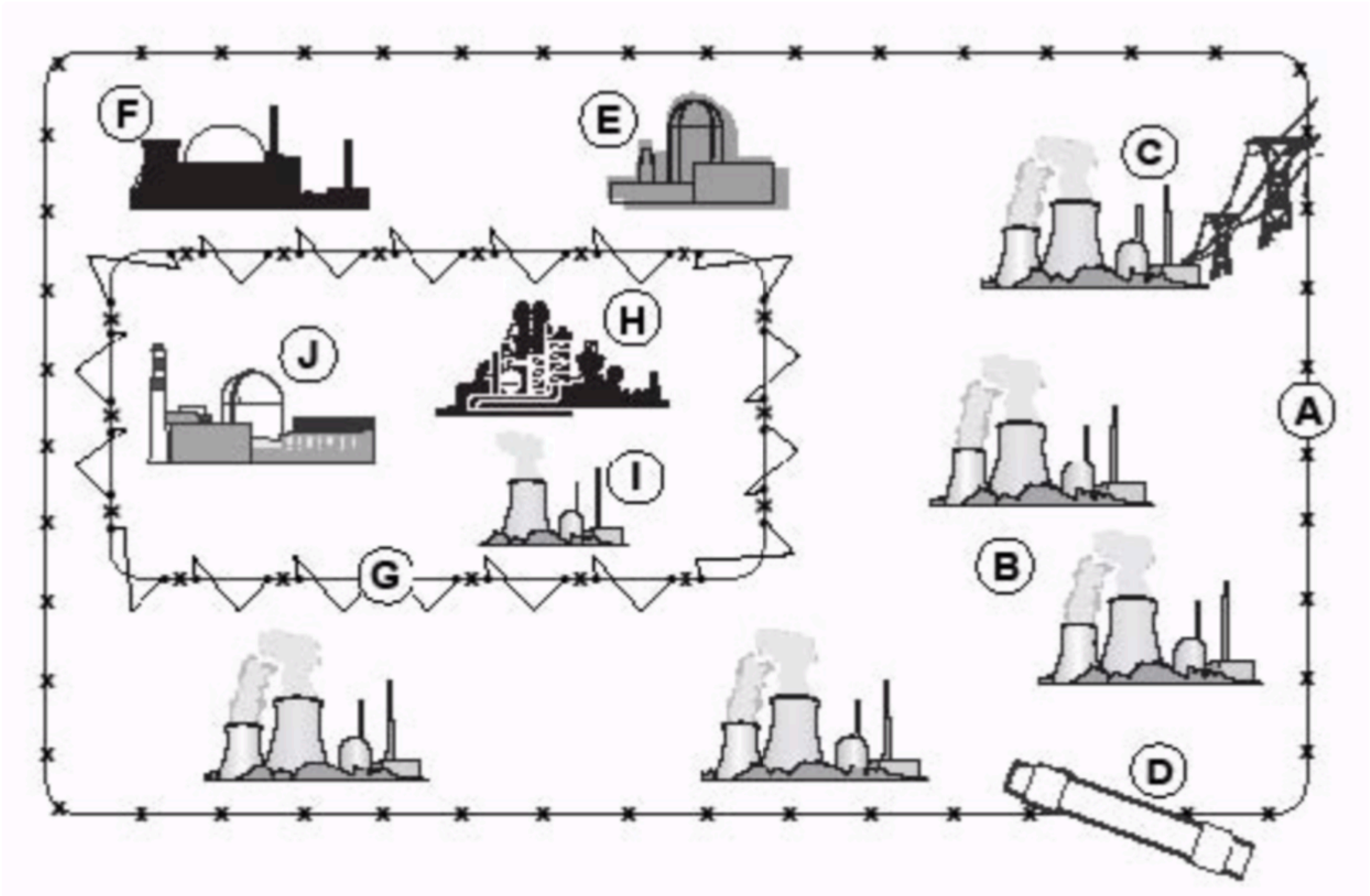

Figure 27: The energy park: A. low security fence; B. 5 thermal 1GWe nuclear reactors, LWRs or more advanced reactors; C. output electricity; D. manufactured fuel pipeline, E. cooling pool for storage of highly radioactive fission products for 300–500 years necessary for them to become inert. This is a time human society can reasonably plan for, unlike the ~ half million years the plutonium 'waste' would continue to be a threat to humankind. A sustainable energy infrastructure should not create a plutonium mine; F. liquid or gaseous fuel factory; G. high security fence, everything with proliferation risk, during the short time before it is diluted or burned, is behind this high security fence; H. separation plant. This separates the material discharged from the reactors (B) into fission products and transuranic elements. Fission products which have commercial value would be separated out and sold, the rest go to storage (E), transuranic elements go to (I); I, the 1GWe integral fast reactor (IFR) or other fast neutron reactor where actinides like plutonium are burned; J. the fusion breeder, producing 1GWe itself and also producing the fuel (ultimately enriched to ~4% $^{233}$U in $^{238}$U) for the 5 thermal nuclear reactors for a total of 7 GWe produced in the energy park.

Note that the energy park uses two levels of security. The low security fence, A, is the just the sort of security that a normal nuclear plant uses. Inside this fence, but outside of the high security fence, there are lots of dangerous materials, but nothing with a proliferation risk. Inside, there is also a high security fence, G, protected by well-trained security personnel, with big guns, and really mean dogs. It must also be protected from penetration by helicopters, perhaps with a steel mesh over it, or perhaps it could be under ground. The material inside is not as dangerous as nuclear weapons, but there will be small quantities of material with proliferation risk. This quantity will be small, because just as soon as the $^{233}$U is produced by the fusion breeder, it will be extracted and diluted with $^{238}$U, or perhaps a mixture of $^{238}$U and thorium. It will be dilute enough in $^{233}$U that it will not be a proliferation risk without isotope separation, which is done in an enormous facility which is difficult to impossible to conceal. The actinides discharged from the thermal reactors, will be initially in a mixture so hot that it must be handled remotely, and only by very specialized equipment. This is immediately brought from the reactors to a spot inside the high security fence. There the actinides are separated from the fission fragments, and as soon as they are separated, they are burned in the fast neutron reactor. The fission fragments will be sent outside the high security fence to the cooling pools, or another standard storage system for intensely radioactive material, but material with no proliferation risk. During a portion of my time at NRL, my office and others were above a cooling pool containing cobalt 60, a highly dangerous and radioactive isotopw. At no time was there any problem. Hence there is neither long time storage, nor long distance travel for any material with proliferation risk. Any material with proliferation risk would be quickly burned or diluted.

The energy park would generate 7 GW of electric power, or some equivalent combination of electric power and liquid or gaseous fuel. The world-wide use of energy parks could generate carbon free power, in an economically and environmentally viable way, and with little or no proliferation risk. They could supply tens of TW at least as far into the future as the dawn of civilization was in the past.

## X: Conclusions

This paper is aware of only 3 means for sustainable power for a midcentury civilization for a world with of 10B people. These are all nuclear options, which tap the vast energy reserve in fertile, instead of the small quantity of available fissile materials. This potential fuel could power a world of 10B people at 40 TW at least as far into the future as the dawn of civilization was in the past. They are fast neutron fission breeding, thermal thorium fission breeding, and fusion breeding (and possibly pure fusion if things work out really well). Hence, we iterate that fusion is an extremely important project for government support. Fusion breeding is less certain than the other two, but as we have seen, it has advantages they do not have. It alone can fuel many thermal reactors which could be 'stranded and out of gas'.

However, the history of fusion research is one of tremendous advances, but also one that has endured many delays and cost overruns. Considering such a history, this author believes that for a very important, but difficult project like fusion, these delays and cost overruns will probably persist. The government bureaucrats supporting fusion should be aware of this. Hence in a worst-case scenario, the government should be able to find a few tens of billions more over a decade or two to support a project that has a reasonable chance of actually working. There will be plenty of fossil and nuclear fuel to get over this speed bump. After all, the government is planning to subsidize solar, wind, and batteries to the tune of tens to hundreds of trillions (Rutchner, Smil, Manheimer 2023A), an expenditure which this author considers to be nearly a total waste. If the government can afford this, it can certainly afford a $10or 20 billion more over a decade or two for a real energy option which has a decent chance of working.

As stated in the body of the text, regarding the different approaches to fusion, nobody's crystal ball is perfectly clear. But some are clearer than others. The conclusion of this paper is that direct drive inertial fusion, powered by an excimer laser, has by far the clearest crystal ball, for both pure fusion and fusion breeding. It seems to this author that this is a perfect way for the United States to recapture the leadership position in fusion it once had. Excimer laser powered fusion and/or fusion breeding is most likely achievable, but only if fusion begins to 'color outside the lines'.

Acknowledgement: This work was not supported by any organization, public or private. Researching and peddling the concept of fusion breeding has been a difficult, but a very satisfying, if lonely endeavor for the author. This research has been difficult to publish in much of the skeptical standard plasma physics literature. The author is indebted to two editors of major American fusion journals which have been willing to consider and accept work on fusion breeding. They are Steven Dean, long time editor of Journal of Fusion Energy, and George Miley, previous editor of Fusion Technology.

This author has published many papers in the highest quality scientific journals. However, while feeling this current paper is of *very* high quality, I decided not to attempt to publish it in a major American scientific journal just yet, if ever. There are three reasons for this. First, it takes at the very least many months and often as long as a year to get it published, and that is if things go smoothly. At my age, that is much too long to wait; not only that I believe it is important to get this material out as soon as possible, so a debate on these proposed options for fusion can get started ASAP. Second, the paper has some very controversial comments about climate and windmills. I am unwilling to remove these comments because they are an important part of the story. However, at the very least they would delay publication further, and more likely render it unpublishable in these journals. Finally, this paper is most likely too long for most of these journals. It has a great deal to say, and I do not feel I can reasonably cut it to for instance half or 2/3 of its length.

This author has attempted several times to publish research skeptical on a climate crisis, and on wind, solar and batteries in several 'normal' American scientific journals, and in each case the articles were rejected immediately, despite being based on solid science. Some were published on what might be considered lesser, 'ignorable' journals. Also, he has attempted on several occasions to publish research on fusion breeding in AIP or APS journals, meeting the same fate. Fortunately for me, several other journals of the same quality, published by Springer, the Cell network, and IEEE have been willing to publish this material on fusion breeding after proper reviewing. In this author's opinion, much of the fusion and scientific establishment, as well as the APS and AIP has been blinded by both the false climate crisis, and by the perfect being the enemy of good enough.

The author is also very lucky to have been aided and encouraged over the years by 9 brilliant scientists and one expert bridge player. These scientists have always been willing to talk with me and point out my missteps, missteps regarding both science and common sense. Without the help of these people, his body of work would have been a much lesser one. Unfortunately, 4 of these people are no longer with us. The first is Martin Lampe. He never personally became involved to the extent of being a coauthor, but until his untimely death, he carefully read every manuscript I wrote on fusion breeding and had many, many very constructive comments. He was my best friend at NRL for many decades. Then there was Dan Meneley, former head of the Canadian nuclear program. He and I met at a conference at Ottawa in 2006 and were in fairly regular contact since then. George Stanford and I never met in real space, but we were certainly friends in cyber space. When I saw his papers on the integral fast reactor, I contacted him by email, and we exchanged many emails on nuclear matters. Paul Parks was a tokamak expert at GA. He kept me abreast on many tokamak matters and believed that breeding was an option that at the very least should certainly not be ignored. He was very skeptical of most of these private 'fusion start ups'. He and I never met in real space but were very familiar with each other in cyber space.

There are 5 still among us, and hopefully will be for a long time. The world certainly could continue to benefit from their wisdom. Ralph Moir has been my tutor on nuclear matters and is an expert on both nuclear and plasma science. He and I were graduate students together at MIT, then went different ways. When I began to take an interest in fusion breeding in ~ 1997, we resumed our contact. I have learned a tremendous amount from him. Jeff Freidberg and I have known each other for a very long time. To my knowledge he is the only other plasma expert to have written a paper on the important impact of 'conservative design rules' for tokamaks. He has been sympathetic to the concept of fusion breeding. At my first seminar on fusion breeding at MIT, he introduced me. In doing so, he mentioned that he had bought season tickets to the Boston Patriots when they were a lousy team and tickets were available and cheap; then Tom

Brady showed up! He wondered if fusion breeding might not have an analogous experience. So far it has not, but this author feels certain that it will; it is virtually inevitable if fusion is to have a future; realistically there is no other option. Nat Fisch is probably about the best theoretical plasma scientist of this era. One certainly gets that impression talking to him, and he has the major awards to prove it. He and I have talked quite a few times about breeding and he has been a real help to me. Jeff Harris and I met briefly in real space at a DoE energy conference in 2009 but have mostly been in touch in cyber space. He has worked with several different groups on tokamaks and stellarators, always communicating with these groups in their language. Then there is Doug Lightfoot, a Canadian nuclear expert, with whom I spent a fair amount of time at that Ottawa meeting along with Dan Meneley. He is an expert on nuclear issues and other energy aspects, and we have been in touch by email since then.

And finally, a nod to Jerry Helms, an expert bridge player who lectures at the national championships, of which we have attended several (these are very welcoming to bridge players at all levels of skill). In my dotage, my wife and I have begun to play bridge in competitive clubs. This can be fun and satisfying, just like studying fusion breeding; but also, can be every bit as frustrating as bashing my head against the wall attempting to convince the establishment of the benefits of fusion breeding. After the publication of my book (Manheimer 2023A), I was resolved to attempt to do my best once more to influence the fusion establishment to both switch a major effort to excimer laser direct drive laser fusion, and to fusion breeding. Of course, there is no hope that I alone could do this, but I had hoped that with variety of media, I could spread the idea in the fusion and larger scientific community. I have done this with a personal letter, a seminar, an essay in the Forum on Physics and Society, and a podcast. However, I wanted to write this major scientific publication on the subject. But so many ideas were floating around randomly in my head; I could not figure out how to organize it, or what its title should be. Then I heard Jerry give a lecture that many aspects of bridge were not covered in bridge textbooks, but there are wide gaps between the precise textbook examples and many actual bridge hands. To deal with this, he said bridge players have to learn to 'color outside the lines'. BINGO! There was my title and organization scheme.

References

Links to the internet are alphabetized ignoring the www, or the https://www